\documentclass[sigconf,10pt]{acmart}
\renewcommand\footnotetextcopyrightpermission[1]{} 
\usepackage{amsfonts,balance,bbold,bm,color,enumitem,hyperref,mathtools,microtype,multicol,multirow,natbib,soul,subfig,tabularx,verbatim,xcolor,xspace}
\usepackage[algo2e,linesnumbered,ruled]{algorithm2e}

\usepackage[intoc]{nomencl}
\makenomenclature

\renewcommand*\nompreamble{\begin{multicols}{2}}
\renewcommand*\nompostamble{\end{multicols}}

\newcommand{\sn}{\textbf{\textit{Leafeon}}\xspace}
\newcommand{\emphasize}[1]{\textbf{\textit{#1}}}
\newcommand{\hj}[1]{\sethlcolor{yellow}\hl{[HJ: #1]}}

\newcommand{\wc}[1]{\sethlcolor{lightgray}\hl{[WC: #1]}}
\newcommand{\revisedtext}[1]{{\color{black} #1}}

\begin{document}

\title{\sn: Towards Accurate, Robust and Low-cost Leaf Water Content Sensing Using mmWave Radar}

\author{Mark Cardamis}
\email{m.cardamis@unsw.edu.au}
\orcid{0000-0001-7896-5038}
\affiliation{%
  \institution{University of New South Wales}
  \streetaddress{K17, Kensington Campus}
  \city{Sydney}
  \state{NSW}
  \country{Australia}
  \postcode{2052}
}

\author{Hong Jia}
\email{hj359@cam.ac.uk}
\orcid{0000-0003-3536-3832}
\affiliation{%
  \institution{University of Cambridge}
  \city{Cambridge}
  \country{UK}
}

\author{Hao Qian}
\email{z5158396@student.unsw.edu.au}
\orcid{0000-0002-7234-7682}
\affiliation{%
  \institution{University of New South Wales}
  \city{Sydney}
  \state{NSW}
  \country{Australia}
  \postcode{2052}
}

\author{Wenyao Chen}
\email{wenyao.chen@unsw.edu.au}
\orcid{0000-0001-7400-132X}
\affiliation{%
  \institution{University of New South Wales}
  \city{Sydney}
  \state{NSW}
  \country{Australia}
}

\author{Yihe Yan}
\email{yihe.yan@student.unsw.edu.au}
\affiliation{%
  \institution{University of New South Wales}
  \city{Sydney}
  \state{NSW}
  \country{Australia}
}

\author{Oula Ghannoum}
\email{o.ghannoum@westernsydney.edu.au}
\orcid{0000-0002-1341-0741}
\affiliation{%
  \institution{Western Sydney University}
  \city{Sydney}
  \state{NSW}
  \country{Australia}
}

\author{Aaron Quigley}
\email{aquigley@acm.org}
\orcid{0000-0002-5274-6889}
\affiliation{%
  \institution{University of New South Wales}
  \city{Sydney}
  \state{NSW}
  \country{Australia}
}

\author{Chun Tung Chou}
\email{c.t.chou@unsw.edu.au}
\orcid{0000-0003-4512-7155}
\affiliation{%
  \institution{University of New South Wales}
  \city{Sydney}
  \state{NSW}
  \country{Australia}
}

\author{Wen Hu}
\email{wen.hu@unsw.edu.au}
\orcid{0000-0002-4076-1811}
\affiliation{%
  \institution{University of New South Wales}
  \city{Sydney}
  \state{NSW}
  \country{Australia}
}

\renewcommand{\shortauthors}{Cardamis et al.}

\begin{abstract}

Plant sensing plays an important role in modern smart agriculture and the farming industry. Remote radio sensing allows for monitoring essential indicators of plant health, such as leaf water content. While recent studies have shown the potential of using millimeter-wave (mmWave) radar for plant sensing, many overlook crucial factors such as leaf structure and surface roughness, which can impact the accuracy of the measurements.


In this paper, we introduce \sn, which leverages mmWave radar to measure leaf water content non-invasively. Utilizing electronic beam steering, multiple leaf perspectives are sent to a custom deep neural network, which discerns unique reflection patterns from subtle antenna variations, ensuring accurate and robust leaf water content estimations.


We implement a prototype of \sn using a Commercial Off-The-Shelf mmWave radar and evaluate its performance with a variety of different leaf types. \sn was trained in-lab using high-resolution destructive leaf measurements, achieving a Mean Absolute Error (MAE) of leaf water content as low as 3.17\% for the Avocado leaf, significantly outperforming the state-of-the-art approaches with an MAE reduction of up to 55.7\%. Furthermore, we conducted experiments on live plants in both indoor and glasshouse experimental farm environments (see Fig.~\ref{fig:teaser}). Our results showed a strong correlation between predicted leaf water content levels and drought events.


\end{abstract}

\maketitle

\section{Introduction}\label{sec:introduction}
The world population is expected to reach 9.7 billion people by 2050~\cite{rayhana_rfid_2021}, increasing the demands on our current food production. The availability of water to plants is vital, as it serves functions such as nutrient transport, plant turgidity, photosynthesis, and as a radiator to cool the plant through transpiration \cite{bloodnick_volumetric_nodate}. However, water unavailability or water stress can occur either seasonally, where soil reserves are depleted, or diurnally when transpiration rates exceed the soil supply rate, challenging the sustainability of food production.

\begin{figure}[t]
    \centering
    \captionsetup[subfloat]{farskip=3pt,captionskip=10pt}
    \subfloat[\revisedtext{Glasshouse experimental farm.}\label{fig:introleaftypeliveplantexperimentsetup}]{\includegraphics[trim={0.5cm 0cm 0.2cm 0.0cm},clip,width=.5\linewidth]{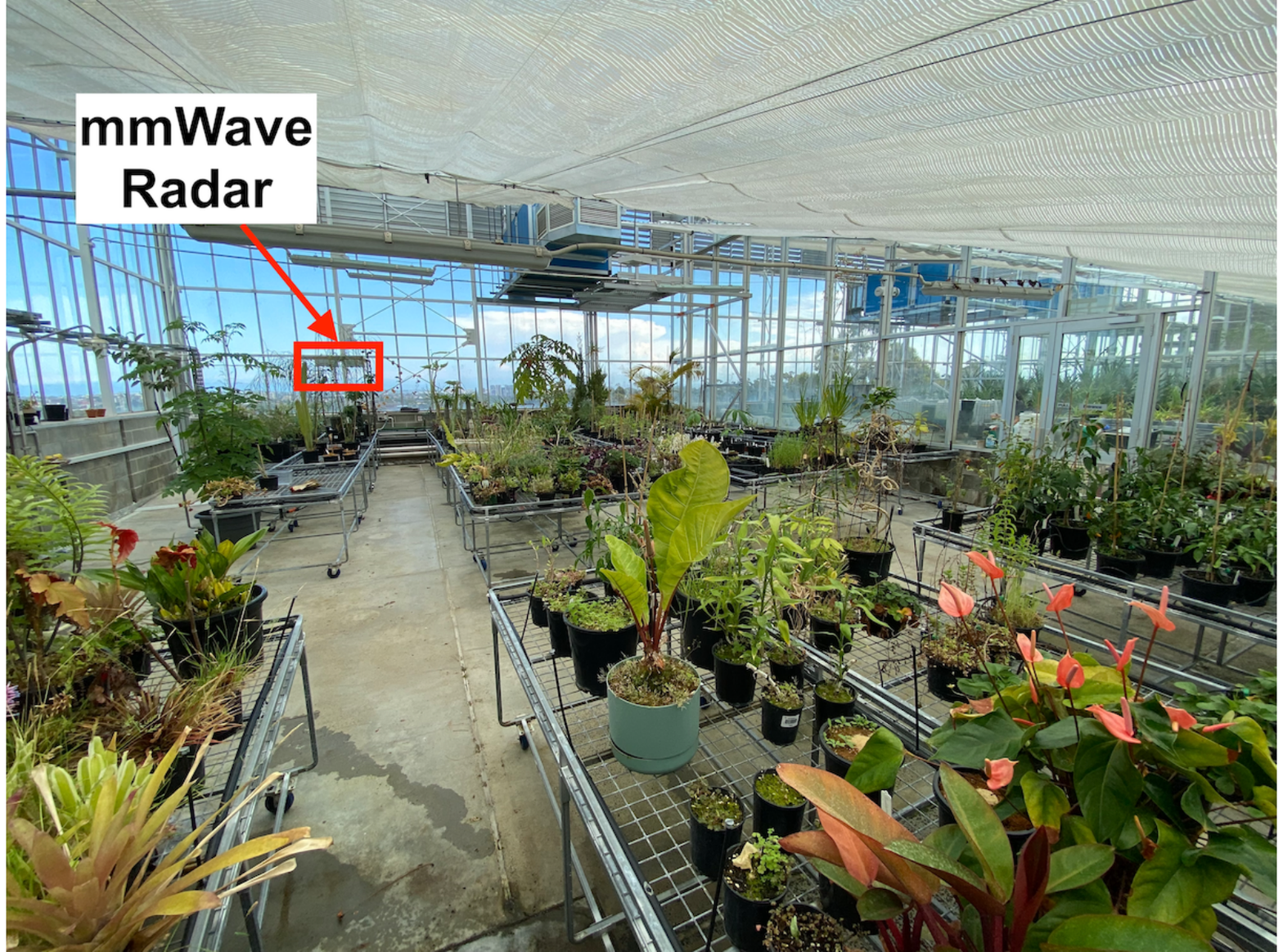}}
    \hspace*{0.5em}
    \subfloat[\revisedtext{Leaf WC Prediction} \label{fig:resultsleaftypeliveoutdoortargetscalerpowertransformer}]{\includegraphics[trim={0.5cm 0.6cm 0.0cm 0cm},clip,width=0.46\linewidth]{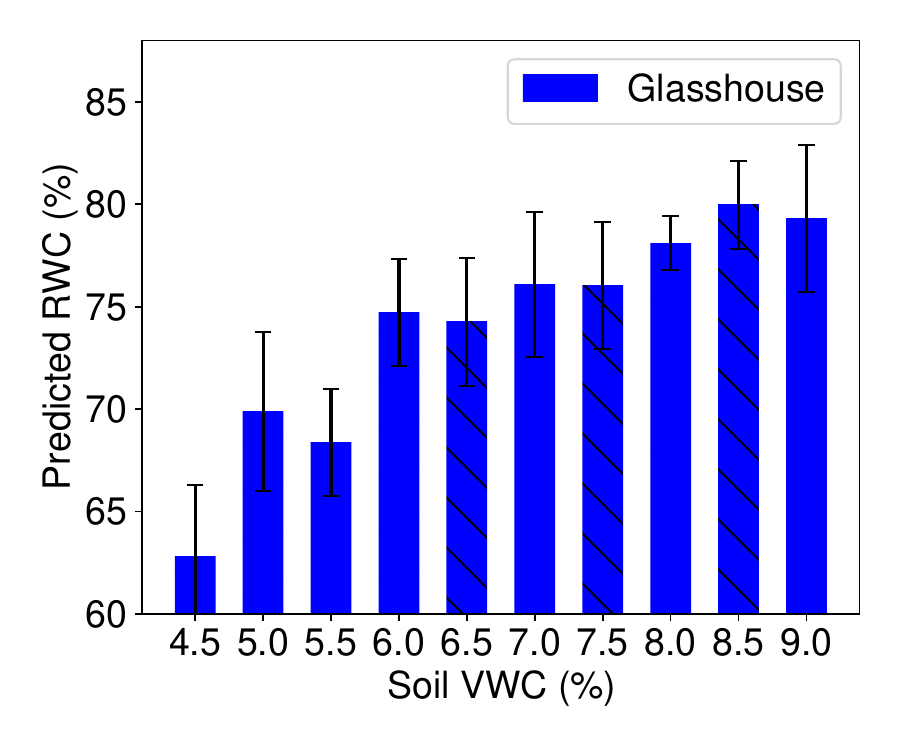}}
    \caption{(a) Measuring leaf water content in Avocado plants in a glasshouse farm (see Fig.~\ref{fig:resultsleaftypeliveplantexperimentsetup} later shows the zoom-in mmWave radar setup). (b) The predicted RWC is strongly correlated to the soil VWC.}
    \vspace{-1em}
    \label{fig:teaser}
\end{figure}

Leaf water potential (LWP) is the most biologically significant signal for identifying water stress according to plant scientists~\cite{jacquemoud_leaf_2019,quemada_remote_2021}. To determine LWP, the relative water content (RWC) of the leaf is commonly calculated, and is inversely related to LWP. RWC is the ratio of the current water content to the maximum water content of the leaf, discussed further in Section~\ref{subsec:goalsMethodologyMetrics}. Thus, by measuring RWC, we can effectively determine the water stress in the plant~\cite{jacquemoud_leaf_2019, govender_review_2009}.

Existing solutions for measuring the RWC face several challenges. First, contact-based commercial products like the Leaf Sensor~\cite{noauthor_leaf_nodate} and PSY1 leaf psychrometer~\cite{noauthor_psy1_nodate} measure leaf water content (WC) by clamping in-situ sensors onto the leaf, which \revisedtext{restricts measurements to a small sample of the entire crop}. Second, infrared thermal imaging~\cite{zhou_assessment_2021,mangus_development_2016,jones_thermal_2003} has been proposed as a non-contact RWC measurement method, although leaf surface temperature does not give a direct measurement of RWC unless accurate evaporative loss rates are known~\cite{atherton_leaf-mounted_2012}. In addition, thermal imaging requires high-end and expensive equipment which is influenced by environmental factors such as sunlight levels or humidity~\cite{atherton_leaf-mounted_2012}.

Recently, there has been a growing interest in real-time, non-contact sensing using radio technologies such as Radio Frequency Identification (RFID)~\cite{zhang_plant_2022}, Long Range Radio (LoRa)~\cite{10.1145/3463526,dang_iotree_2022}, WiFi~\cite{peden_model_2021} and millimeter-wave (mmWave) radar~\cite{santos_potential_2021}. Radio waves strongly interact with water~\cite{jiang_electromagnetic_2011}, making it possible to measure changes in leaf WC. Compared to RFID and WiFi, mmWave provides much larger radio bandwidths~\cite{peden_model_2021,10.1145/3534601}, improving the accuracy of leaf WC estimation. A larger bandwidth enables more data to be collected by the system, yielding more detailed spatial observations. Despite some promising preliminary results using mmWave for RWC measurements~\cite{hoog_60_2022}, previous works only conducted in-lab experiments using detached leaves \revisedtext{and didn't test on live plants}. They limit the accuracy of the system by \revisedtext{concentrating exclusively on the reflective radio signal, assuming the leaves are smooth and flat surfaces. They neglect to investigate how surface roughness and internal leaf structure cause different backscattered energy at different points of the leaf}.
In fact, the leaf's surface roughness has an impact on the reflected signal at mmWave frequencies, additionally, the internal structure of the leaf must be considered because the reflected signal is a result of both surface and internal volumetric scattering~\cite{jacquemoud_leaf_2019}. As a result, there is currently a lack of a non-invasive and accurate system for measuring leaf WC.

In this study, we present \sn, a novel non-contact system that considers the leaf's surface and volumetric scattering effect to improve the accuracy of the WC estimation. Our system explores beam steering and Angle of Arrival (AoA) properties of commercially available mmWave radar technology and examines their correlation with leaf WC. We then propose a method that takes into account both the radio's signal strength and location information to improve measurement accuracy. Following, we tailored a novel deep learning (DL) model in \sn to learn a representation between beam steering, AoA and leaf WC measurements. By using electronic beam steering, \sn takes advantage of the unique differences in radio wave propagation paths by examining both surface and volumetric scattering, which has not been considered in previous works. The DL model then automatically selects the most relevant features to predict reliable WC measurements, combining information from radio signals received at various incident angles to infer accurate WC predictions.

We implement \sn with a COTS mmWave radar and demonstrate its performance by evaluating it in-lab using real-world data collected from three types of leaves. Our in-lab evaluations show that \sn can achieve MAE accuracy of as low as 3.17\% when predicting different WC levels. The in-lab model was then tested on live plants in both indoor and glasshouse environments, where the data strongly correlated with the drought events. In summary, we make the following contributions:
\begin{itemize}
  \item We design a general working principle using transmit beam steering and receiver AoA beamforming for accurate and robust leaf WC monitoring using radio, which investigates the wave propagation in multilayered leaf structures to improve the WC prediction. We demonstrate that the reflected radar signal depends on both the water levels inside the leaf and the radio wave incident angle.
  \item Based on the above analysis, we propose \sn, which uses a DL model to fuse the information from radio transmit beam steering and receiver beamforming to increase the prediction accuracy and robustness. To the best of our knowledge, we are the first to use radar beam steering and AoA on a dynamic organic surface, such as a leaf, to monitor its physical properties. 
  In comparison, previous works \cite{wu_msense_2020,liang_fg-liquid_2021} on using radar for material and liquid identification did not consider beam steering properties.  
  \item We conduct extensive in-lab experiments with different types of \revisedtext{detached} leaves to show that \sn can achieve high accuracy when predicting WC from reflected radar signals. \sn outperforms the state-of-the-art (SOTA) ~\cite{hoog_60_2022} with a MAE of 3.17\% for the Avocado leaf, a 55.7\% improvement.
  \item We evaluate the system on live plants in both indoor and glasshouse experimental farm environments, showing a strong relationship between the predicted leaf WC and drought events, \revisedtext{highlighting the feasibility of using detached leaves for training and live plants for testing, see Fig.~\ref{fig:teaser}.}
\end{itemize}

\begin{figure}[t]
  \centering
    \subfloat[Plant radar reflections \label{fig:backgroundradarpowervswatera}]{\includegraphics[width=0.38\linewidth]{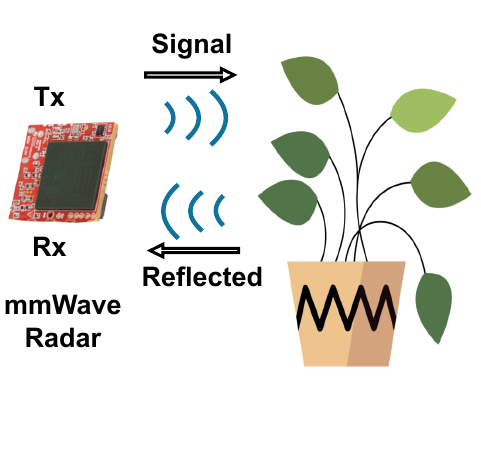}}
    \hspace*{1.0em}
    \subfloat[RSS vs Distance \label{fig:backgroundradarpowervswaterb}]{\includegraphics[width=0.55\linewidth]{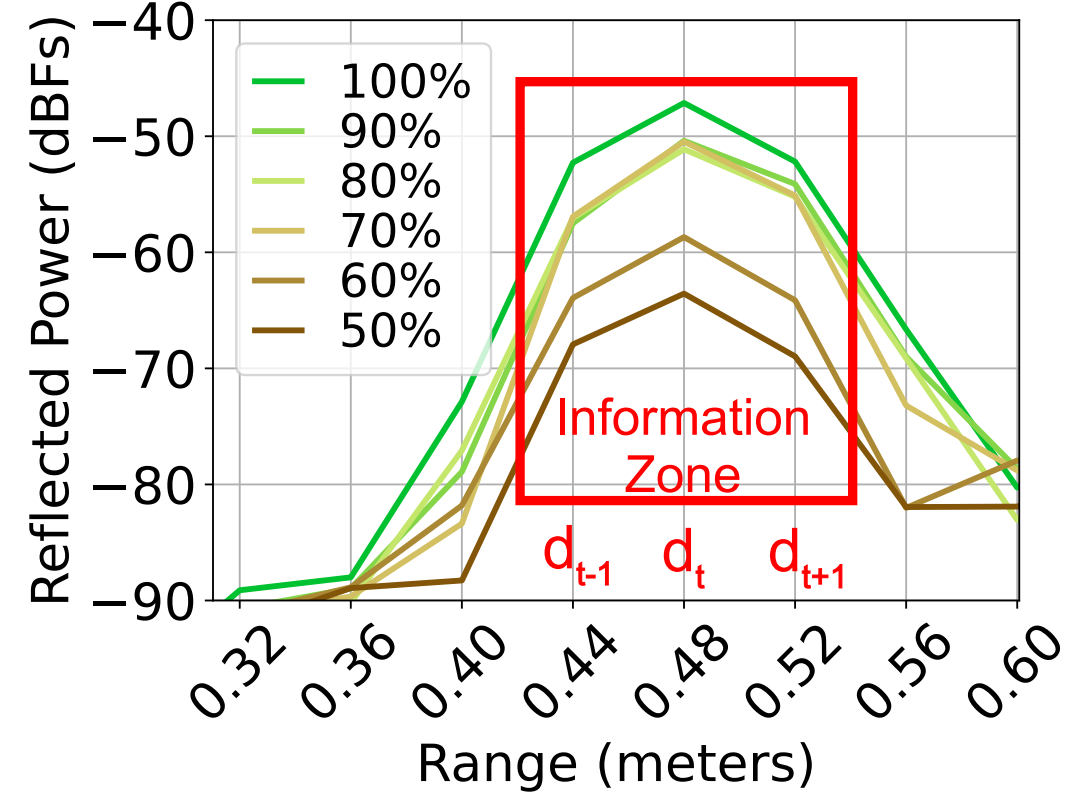}}
  \caption{mmWave signals reflected by leaves with different WC levels (50-100\% in (b))}
  \label{fig:backgroundradarpowervswater}
  \Description{The graphs show the reflected power (dBFS) for various leaf WC levels}
\end{figure}

\section{Preliminary Study and Challenges}\label{sec:background}
\subsection{Radar Overview}\label{subsec:radaroverview}

Radar operates by transmitting an electromagnetic (EM) wave and measuring the received signal after it reflects off an object (see Figure~\ref{fig:backgroundradarpowervswatera} for an example). 
In the context of biological materials, WC is one of the major components in the reflection coefficient \cite{chang_permittivity_2005}, which affects the radar's received signal (see Figure \ref{fig:backgroundradarpowervswaterb} for an example). Specifically, the Received Signal Strength (RSS) represented by amplitude $A_d$ \nomenclature{$A_d$}{Radar signal amplitude received} can be calculated using a modified Friis transmission formula\footnote{Here, we use signal amplitude instead of power, while traditional Friis Transmission Formula is defined as $P_r = P_t G_t G_r \lambda^2 / (4 \pi D)^2$, where $P_r$ and $P_t$ are the receive and the transmit power respectively, $G_t$ and $G_r$ are the power gain of the transmitting and receiving antennas respectively and $D$ is the propagation distance.}: 
\begin{equation}\label{eqn:friis_transmission_reflection}
  A_d = A_o g_t g_r \frac{\lambda}{4 \pi (2d_t)} \cdot{r},
\end{equation}
where $A_o$ \nomenclature{$A_o$}{Radar signal amplitude transmitted} is the amplitude of the transmitting signal, $g_t$ \nomenclature{$g_t$}{Square-root of radar transmitter antenna gain} and $g_r$ \nomenclature{$g_r$}{Square-root of radar receiver antenna gain} are the gains of the Transmitter (Tx) and Receiver (Rx), respectively, $d_t$ \nomenclature{$d_t$}{Distance to the radar target} is the distance to the target, $\lambda$ is the wavelength, and $r$ \nomenclature{$r$}{Material reflection coefficient} is the reflection coefficient of the reflected object, which will be discussed in detail in Section~\ref{subsec:reflectionCoefficient}. This equation applies to a typical monostatic radar, \textit{i.e.} the Tx and Rx are co-located.

As the received radar signal {$A_d$} is processed using Fast Fourier Transform (FFT) into discrete bins \cite{10.1145/3447993.3483251}, the leaf WC information may fall into more than one bin (i.e., distance interval). To include all range bins that contain leaf water content information, \sn explores the reflected radar signal from three range bins ($d_{t-1}$ \nomenclature{$d_{t-1}$}{Distance of the range bin before the target}, $d_{t}$, and $d_{t+1}$\nomenclature{$d_{t+1}$}{Distance of the range bin after the target}), which form the \textbf{leaf information zone}, highlighted in Figure~\ref{fig:backgroundradarpowervswaterb}. The reflected radar power is correlated to leaf WC and can be calculated using the \textit{reflection coefficient}, which will be discussed in detail in Section~\ref{subsec:reflectionCoefficient}.

\begin{figure}[t]
  \centering
  \includegraphics[width=\linewidth]{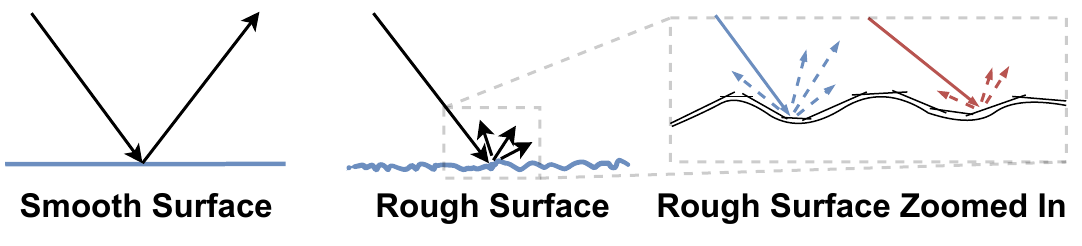}
  \caption{Radio energy scattered from a smooth and a rough surface, respectively.}
  \label{fig:backgroundradarscatteringsurface}
  \Description{The directional pattern of the energy scattered by both a smooth and rough surface. For a rough surface, the scattering profile is different across the leaf's surface.}
\end{figure}

\subsection{Reflection Coefficient}\label{subsec:reflectionCoefficient}

When the radar signal hits an object, e.g., a leaf surface, the material permittivity governs how much of the signal will be reflected~\cite{chang_permittivity_2005} and can be calculated with the Fresnel Equation: 
\begin{equation}\label{eqn:reflection_refraction}
  r = \frac{n - 1}{n + 1}, 
\end{equation}
where $n$ \nomenclature{$n$}{Refractive index of the material} is the refractive index of the material. Furthermore, the refractive index, $n$, of a material may be calculated using the complex permittivity of the material: 
\begin{equation}\label{eqn:refraction_permittivity}
  n = \sqrt{\frac{1}{2}\sqrt{({e_r}^{'})^2 + ({e_r}^{''})^2} + {e_r}^{'}},
\end{equation}
where ${e_r}^{'}$ is the real part of the relative permittivity, and ${e_r}^{''}$ is the imaginary part of the relative permittivity.

Therefore, the leaf refractive index $n$ may be calculated by Eq.~(\ref{eqn:reflection_refraction}) via the reflection coefficient $r$, which, in turn, may be calculated by Eq.~(\ref{eqn:friis_transmission_reflection}) based on the measured RSS $A_d$ values in the Rx antennas of a radar. 

One of the limitations of prior work is that it assumes that the leaf surface is smooth and specular reflection occurs~\cite{hoog_60_2022}. However, since the wavelength of mmWave radio is very small (i.e., in the millimeter range), the diffusing and scattering effects of radio waves created by the rough surface of a leaf will have a significant impact on the measured reflected signal $A_d$ and therefore cannot be ignored. 

\textbf{\emph{Challenge 1}}. Hence, measuring the water content from a rough leaf surface is challenging as it is difficult to isolate from the surface scattering effects in $A_d$.

\subsection{Radio Wave Scattering}\label{subsec:radioWaveScattering}


A smooth surface causes specular reflection of the incident wave~\cite{chen_mmcamera_2022} (see Figure~\ref{fig:backgroundradarscatteringsurface} \textit{left}), where the incident angle reflects the backscattered energy away from the radar. A rough surface ($\Delta height > \lambda/32$~\cite{noauthor_34_nodate}), will scatter the incoming wave, with the backscattered energy traveling in all directions (see Figure~\ref{fig:backgroundradarscatteringsurface} \textit{middle}). For an uneven rough surface, the backscattered energy differs across the surface due to the varying scattering points (see Figure~\ref{fig:backgroundradarscatteringsurface} \textit{Rough Surface Zoomed In} for illustration). Thus, in order to estimate the WC in a leaf accurately, the scattering effect needs to be considered.

\begin{figure}[t]
  \vspace{-1em}
  \centering
  \subfloat[Leaf structure \label{fig:methodleafstructurea}]
  {\includegraphics[width=0.48\linewidth]{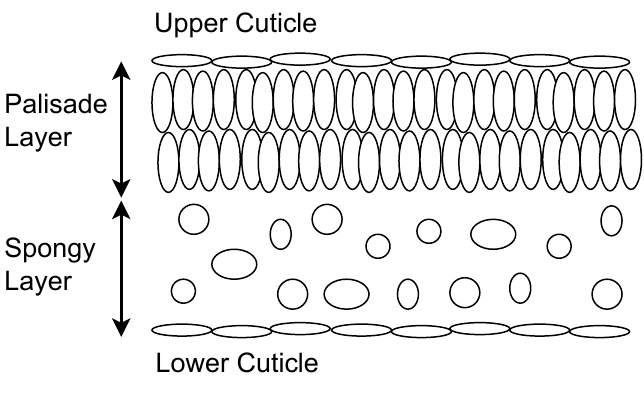}}
  \hspace*{1.0em}
  \subfloat[Wave propagation through a leaf\label{fig:methodleafstructureb}]{\includegraphics[width=0.4\linewidth]{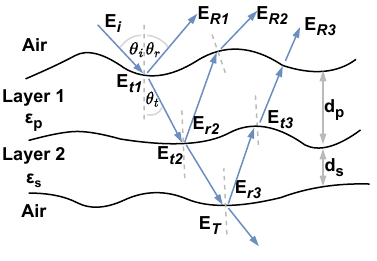}}
  \caption{The structure of a leaf visible to a mmWave radio signal}
  \label{fig:methodleafstructure}
  \Description{Two different layers of cells can be distinguished at mmWave frequencies in a plant leaf. Each layer has distinct dielectric properties, due to their varying water content, creating multiple reflection boundaries for the wave.}
  \vspace{-1em}
\end{figure}

\subsection{mmWave Interaction in a Leaf Structure}\label{itm:background_leaf_structure}

The leaf structure is not a homogeneous medium but comprised of multiple layers with different WC levels. At mmWave frequencies, the wavelength may be small enough to distinguish two different layers in the leaf, each with its own dielectric profile \cite{sarabandi_millimeter_1990}, shown in Figure~\ref{fig:methodleafstructurea}. Specifically, the WC is higher in the upper layer, i.e., Palisade layer, of the leaf than the lower layer, i.e., Spongy Layer, at a ratio of 4:1~\cite{sarabandi_millimeter_1990}.

When a target consists of multiple layers with distinct dielectric properties, internal reflections occur at each boundary and affect the scattering process, shown in Figure~\ref{fig:methodleafstructureb}. Specifically, \emphasize{Surface scattering} refers to the reflected signal from the uppermost interface, $E_{R1}$, and is governed by the permittivity, the roughness of the surface, and the incident angle \cite{ulaby_handbook_2019}. It is the dominant scattering effect at higher WC levels due to the higher reflection coefficient derived from the permittivity. \emphasize{Volumetric scattering} refers to the reflected signal coming from the internal layers, $E_{R2}$ + $E_{R3}$ + .., and is less dependent on the surface roughness because it operates by the principle of multiple scattering. Volumetric scattering is more dominant at lower WC levels because more of the radio wave energy penetrates the uppermost interface~\cite{eliran_empirical_2013}. The reason that \textit{Surface scattering} is \revisedtext{more} dependent on the incident angle \revisedtext{than} \textit{Volumetric scattering} is explained next using Snell's Law of Refraction.

\begin{figure*}[t]
  \centering
  \includegraphics[width=\textwidth]{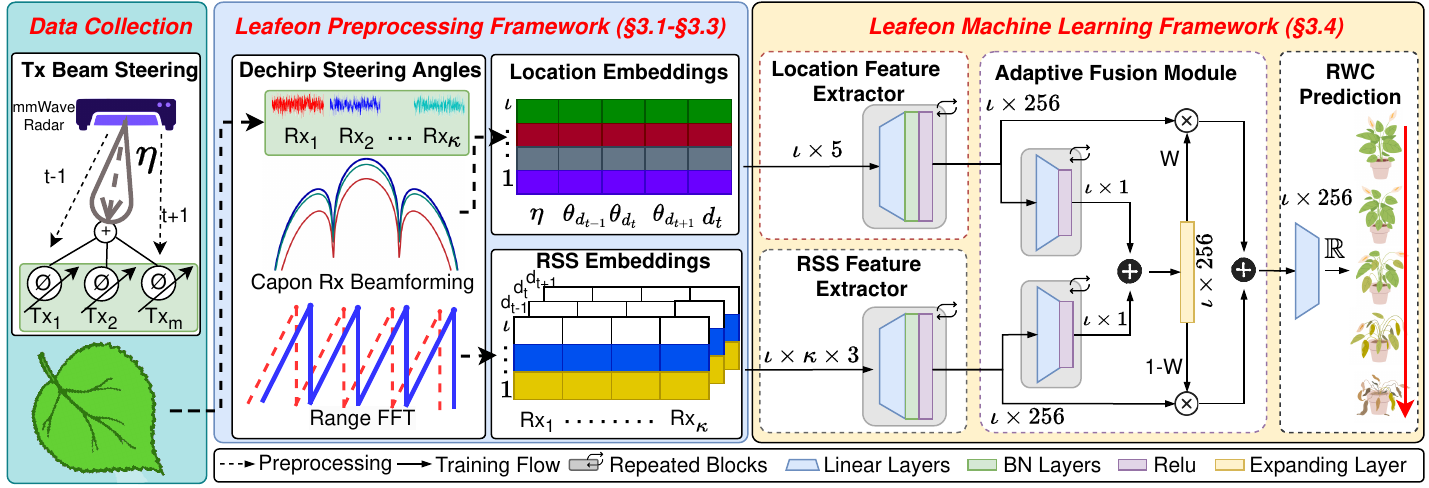}
  \caption{LM-Net network uses processed radar signals as its input, which contain information from $\kappa$ Rx antennas. The signals are divided into Location and RSS input vectors, which are then passed into feature extraction modules. The feature extraction outputs are then fused using adaptive weights to the LM-Net output, which predicts the WC of a leaf.}
  \label{fig:methodoverviewsystemworkflow}
  \Description{A block diagram representing the system workflow of \sn}
  \vspace{-1em}
\end{figure*}

\subsubsection{Snell's Law of Refraction} 

When a radio wave travels from one medium to another, the wave will bend towards the medium that has the lower velocity of propagation, with the relationship of the incident angle $\theta_i$ to the refracted angle $\theta_t$ defined as:
\begin{equation}\label{eqn:snells_law}
 n_i \cdot sin(\theta_i) = n_t \cdot sin(\theta_t),
\end{equation}
where $n_i$ \nomenclature{$n_i$}{Refractive index of the first material} and $n_t$ \nomenclature{$n_t$}{Refractive index of the second material} are the refractive indices of the two media, respectively. In the context of leaf WC measurement with radar (see Figure~\ref{fig:backgroundradarpowervswatera}), the refractive index of air $n_i$ is 1 and the refractive index of the leaf $n_t$ may be calculated by Eq.~(\ref{eqn:friis_transmission_reflection}) and Eq.~(\ref{eqn:reflection_refraction}) based on the measured RSS $A_d$ values in the Rx antennas as discussed earlier. Furthermore, Eq.~(\ref{eqn:snells_law}) also shows that the AoA of the received radio signal may also have an impact on the estimation of $n_t$ in the rough (leaf) surface, which inspires us to consider AoA as an input (model feature) in \sn.

Since a leaf structure has multiple layers (see Figure \ref{fig:methodleafstructure}), the velocity difference between the layers will govern how the wave refracts. In the context of leaf WC measurement with radar, the velocity change from air to Layer 1 (with refractive index $n_p$) \nomenclature{$n_p$}{Refractive index of the leaf palisade layer} is significantly greater than that from Layer 1 to Layer 2 (with refractive index $n_s$)\nomenclature{$n_s$}{Refractive index of the leaf spongy layer}. This is because the Palisade layer has four times higher water content than the Spongy layer, resulting in a refractive index of the Palisade layer $n_p$ that is higher than the Spongy layer's refractive index $n_s$ ~\cite{sarabandi_millimeter_1990}. This is the reason that \textit{surface scattering}, which occurs at the uppermost interface, \textit{i.e.} air to Layer 1, is more dependent on the incident angle than \textit{volumetric scattering}.

\textbf{\emph{Challenge 2}}. As such, as the surface of the leaf is not smooth, effectively extracting information on the surface and volumetric scattering from different incident angles is challenging.

\begin{figure}[htbp]
  \centering
  \subfloat[Tx beam across leaf \label{fig:methodtxbeamsteeringa}]{\includegraphics[height=0.38\linewidth]{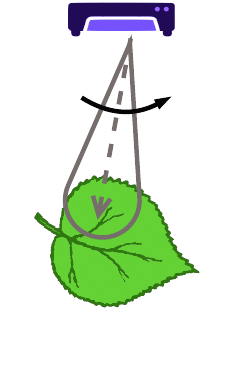}}
  \subfloat[Steering beam using Tx phase shift $\phi$ \label{fig:methodtxbeamsteeringb}]{\includegraphics[height=0.4\linewidth]{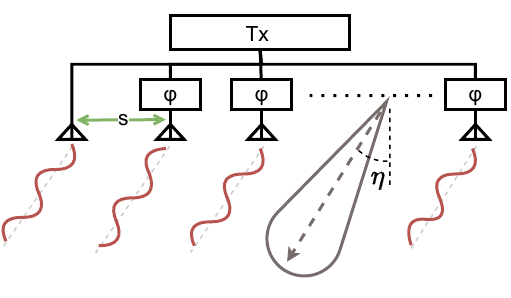}}
  \vspace{-0.5\baselineskip}
  \caption{Using Tx beam steering to control the direction of the radar beam}
  \label{fig:methodtxbeamsteering}
  \Description{A physical representation of Tx beam steering where the radar beam direction is steered in the $\eta$ direction by configuring the phase shift $\phi$, calculated using Equation \ref{eqn:aoaphaseomega}. The phase offset is based on the extra distance a wave must travel with respect to the first Tx so the wave constructively interferes in the $\eta$ direction.}
  \vspace{-1em}
\end{figure}

\section{\sn Overview}\label{sec:methodology}

To address the discussed challenges, in this section we discuss \sn, a framework to estimate leaf WC using radar reflection signal measurements, with an overview shown in Figure \ref{fig:methodoverviewsystemworkflow}. First, we discuss using Tx beam steering and Rx beamforming to exploit the physical signal interactions to extract fine-grained WC changes in the leaf. We then design LM-Net, a robust DL model that can learn a generalized representation of the leaf WC by mapping the radar Rx signal measurements, i.e., RSS ($A_d$) and AoA, from different Tx beam steering angles, to the leaf WC levels.


\subsection{Transmit Beam Steering}

Tx beam steering refers to the ability to control the direction in which a radio signal is transmitted~\cite{feng_rflens_2021, yang_side-lobe_2022}, shown in Figure~\ref{fig:methodtxbeamsteeringa}. Figure~\ref{fig:methodtxbeamsteeringb} shows that beam steering may be achieved by adjusting the phase of the signal $\phi$ \nomenclature{$\phi$}{Radar signal phase delay in transmitter array} in the Tx antenna path according to:
\begin{equation}\label{eqn:aoaphaseomega}
    \phi_m = 2 \pi \frac{s_{m}}{\lambda} \sin(\eta),
\end{equation}
where $s_m$ \nomenclature{$s$}{Spacing between antenna elements on the radar} is the space (distance) between antennas Tx$_m$ and Tx$_1$, and $\eta$ \nomenclature{$\eta$}{Radar transmitter beam steering direction} is the desired angle of the Tx radar radio wave. Note that $s_1$ = 0.

\begin{figure}[t]
  \centering
  \subfloat[Reflected power vs Steering angle for two RWC levels \label{fig:resultsavocadoleafair14steeringanglea}]{\includegraphics[width=0.49\linewidth]{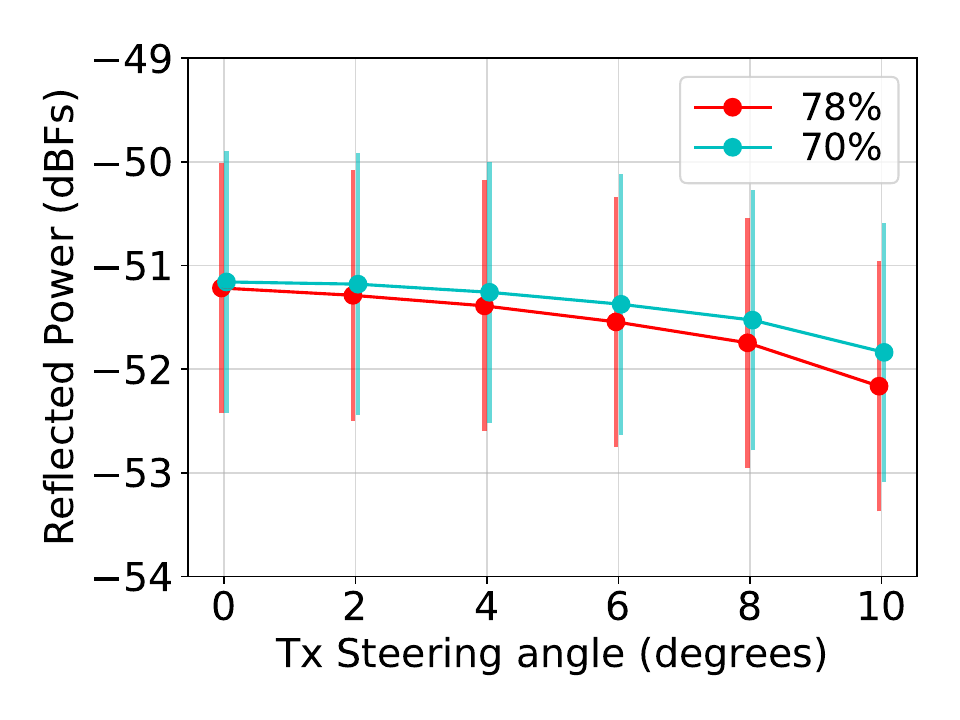}}
  \hfill
  \subfloat[dB delta for high (8º/10º) and low (0º/2º) steering angles \label{fig:resultsavocadoleafair14steeringangleb}]{\includegraphics[width=0.48\linewidth]{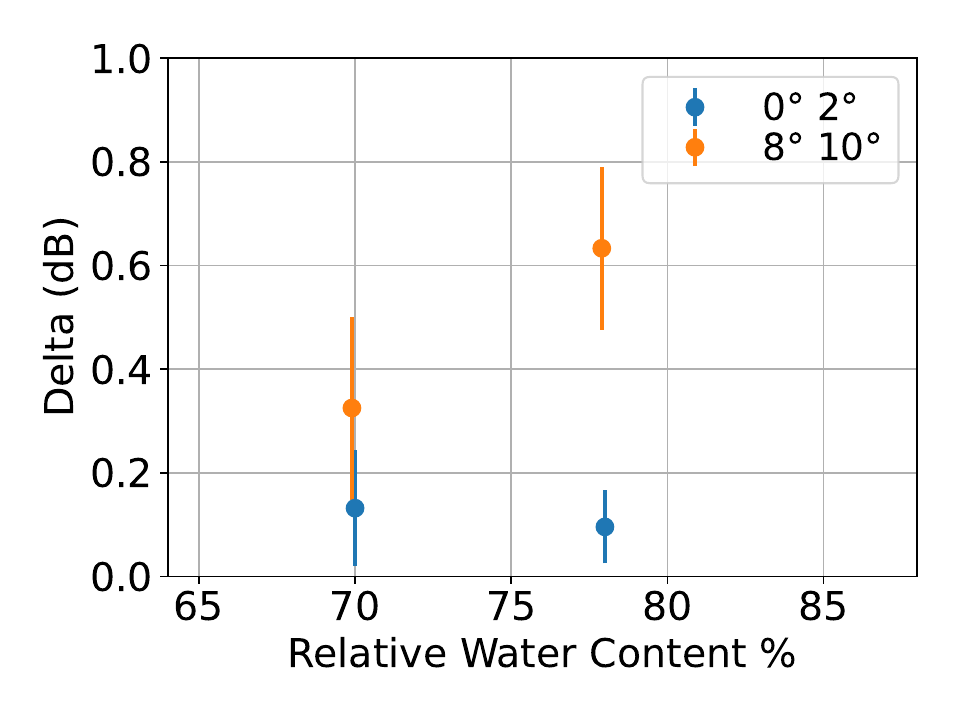}}
  \caption{RSS for the radio waves from different steering angles.}
  \label{fig:resultsavocadoleafair14steeringangle}
  \Description{Multiple measurements of the leaf were taken using Tx beam steering, creating various incident angles as the radar scanned across the leaf face. The difference in dB between steering angles 8º/10º and 0º/2º were plotted for two RWC levels.}
\end{figure}

By steering the radar beam in a specific direction $\eta$, we may control the leaf incident angle ($\theta_i$)\nomenclature{$\theta_i$}{Radar incident angle from air-leaf interface}, shown previously in Figure~\ref{fig:methodleafstructureb}, which will affect the RSS ($A_d$) and refracted angle ($\theta_t$) \nomenclature{$\theta_t$}{Radar refracted angle from air-leaf interface} of the wave. Since the radio waves from different incident angles will have unique scattering characteristics, affected by the WC levels and internal structure of the leaf, as discussed in Section~\ref{itm:background_leaf_structure} earlier, combining the radio wave information from multiple incoming incident angles may be able to produce a better leaf WC level estimation than that from a single (e.g., normal incident angle) beam only.

As surface scattering is more dependent on permittivity $e_r$ \nomenclature{$e_r$}{Material relative permittivity} and leaf incident angle $\theta_i$ than volumetric scattering, the reflected radar signal is more sensitive to the incident angle at higher WC levels. Figure~\ref{fig:resultsavocadoleafair14steeringanglea} shows the RSS for the radio waves from different steering angles (from 0º and 10º) at two different Relative WC (RWC) levels (70\% and 78\%) with the error bars representing the standard deviation over different experiments. The differences in RSS between the two RWC levels increase with higher Tx steering angles. For example, when the Tx steering angle is 0º, the difference of RSS for the two RWC levels is less than 0.1 dBFs. When the Tx steering angle is 10º, the difference of RSS for the two RWC levels increases to approximately 0.4 dBFs, representing a four-fold increase.

At higher WC levels, the radar incident angle has a larger effect on the reflected power \cite{eliran_empirical_2013}. Figure~\ref{fig:resultsavocadoleafair14steeringangleb} demonstrates this by plotting the difference between the radio waves, over different experiments, from high incident steering angles (i.e., 8º and 10º) and those from low incident steering angles (i.e., 0º and 2º), but with the same angle interval (i.e., 2º). We can see that the RSS difference between the waves from 8º and 10º steering angles is larger at the 78\% RWC level than that at the 70\% RWC level, while the difference between the waves from the 0º and 2º Tx steering angles results barely changed. Therefore, we may use the RSS difference between the radio waves from different incident angles as features to estimate leaf WC levels. 

\begin{figure}[t]
  \centering
  \subfloat[][Reflected power vs Steering angle for 100\% RWC \label{fig:methodavocadoleaf7dryinga}]
  {\includegraphics[width=0.49\linewidth]{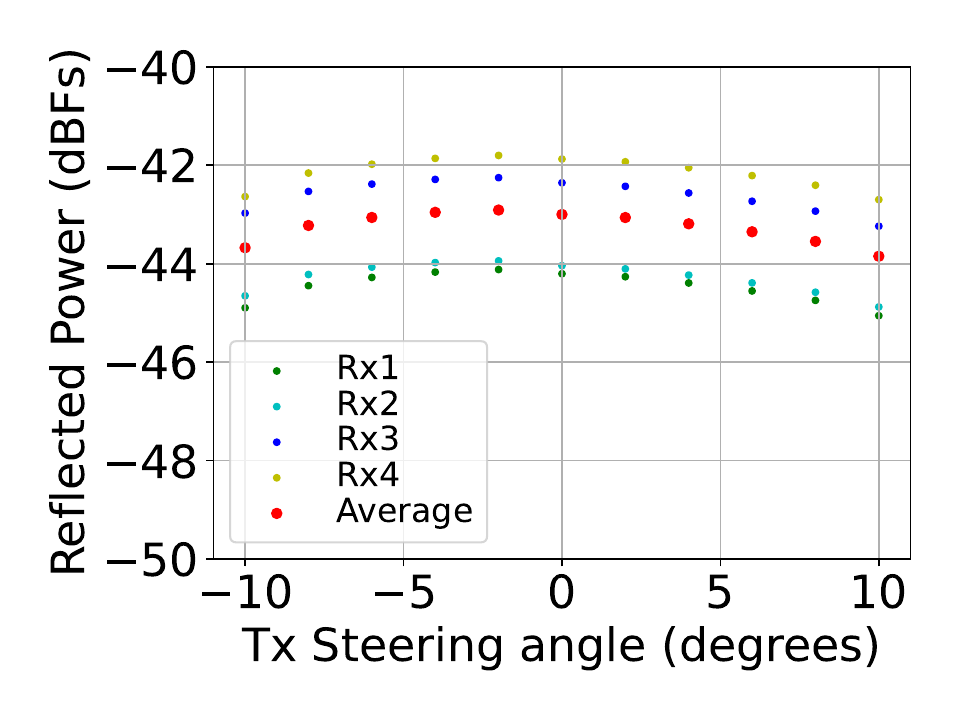}}   
  \hfill
  \subfloat[][Reflected power vs Steering angle for 70\% RWC \label{fig:methodavocadoleaf7dryingb}]{\includegraphics[width=0.49\linewidth]{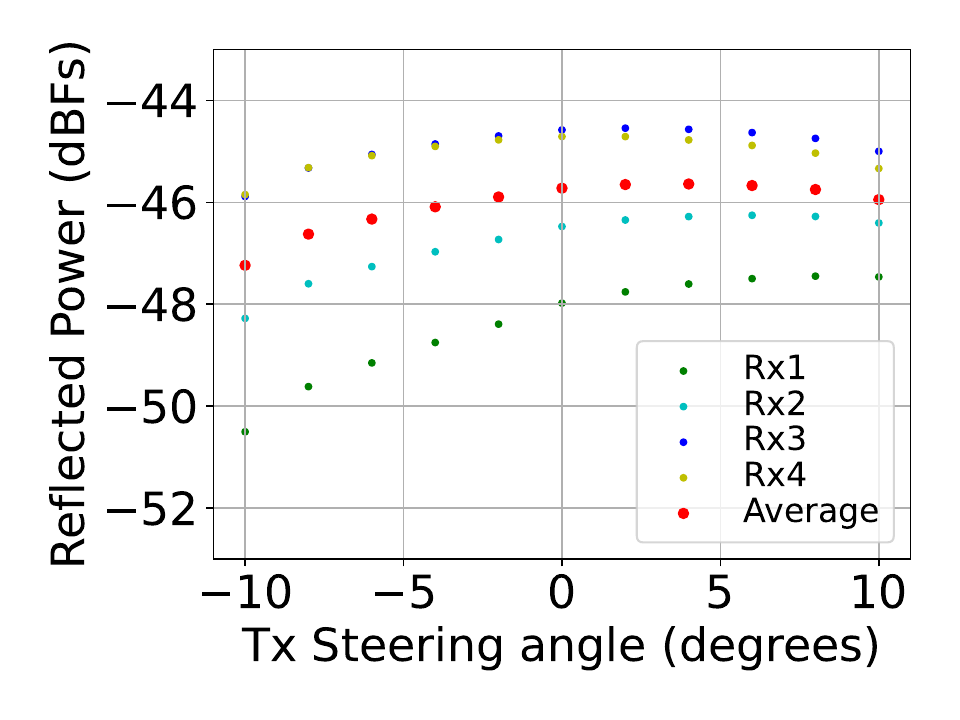}} 
  \caption{Reflected Power for various Tx Steering angles using two different leaf RWC levels}
  \label{fig:methodavocadoleaf7drying}
  \Description{Reflected Power for various Tx Steering angles using two different leaf RWC levels}
\end{figure}

Furthermore, Figure~\ref{fig:methodavocadoleaf7drying} shows the distributions of RSS among Rx antennas from various Tx steering angles vary for different leaf RWC levels. For example, when the RWC is 100\%, the RSS peaks for the radio waves at Tx steering angles around -3º (see Figure~\ref{fig:methodavocadoleaf7dryinga}). When the RWC is 70\%, the RSS peaks for the radio waves at Tx steering angles around 3º (see 
Figure~\ref{fig:methodavocadoleaf7dryingb}). Therefore, another useful aspect for fine-grained leaf WC estimation may be the distributions of RSS from various Tx steering angles.

\subsection{Leaf Scattering Modeling}
In order to better understand the relative contribution of surface and volumetric scattering in a leaf, we create a mmWave scattering model, extending previous works~\cite{sarabandi_millimeter_1990}. The model simulates the Radar Cross Section (RCS) as a function of radar incident angle for both High (100\% RWC) and Low (50\% RWC) moisture levels. The leaf is modeled as a rectangular sheet with two layers where the top and bottom layers correspond to, respectively, the palisade and spongy layers. Each layer has its own permittivity and thickness, which are functions of the moisture level. The model assumes that the radio wave hits the palisade layer at an angle of incidence of $\theta$, with dominant reflections at the surface of palisade layer and at the interface between the two layers. Eq.~(14) in ~\cite{sarabandi_millimeter_1990} gives an expression on the RCS $\sigma(\theta_s,\theta)$ when the incident angle is $\theta$ and the scattering angle is $\theta_s$. Since we use a monostatic radar where the transmitter and receiver are co-located, the RCS of interest is $\sigma(\theta_s,\theta)$ where $\theta_s = \theta$. 

\begin{figure}[t]
  \centering
  \vspace{-1.0\baselineskip}
  \subfloat[Reflections from each leaf layer at high RWC\label{fig:methodrcslayersrwc100}]{\includegraphics[width=0.48\linewidth]{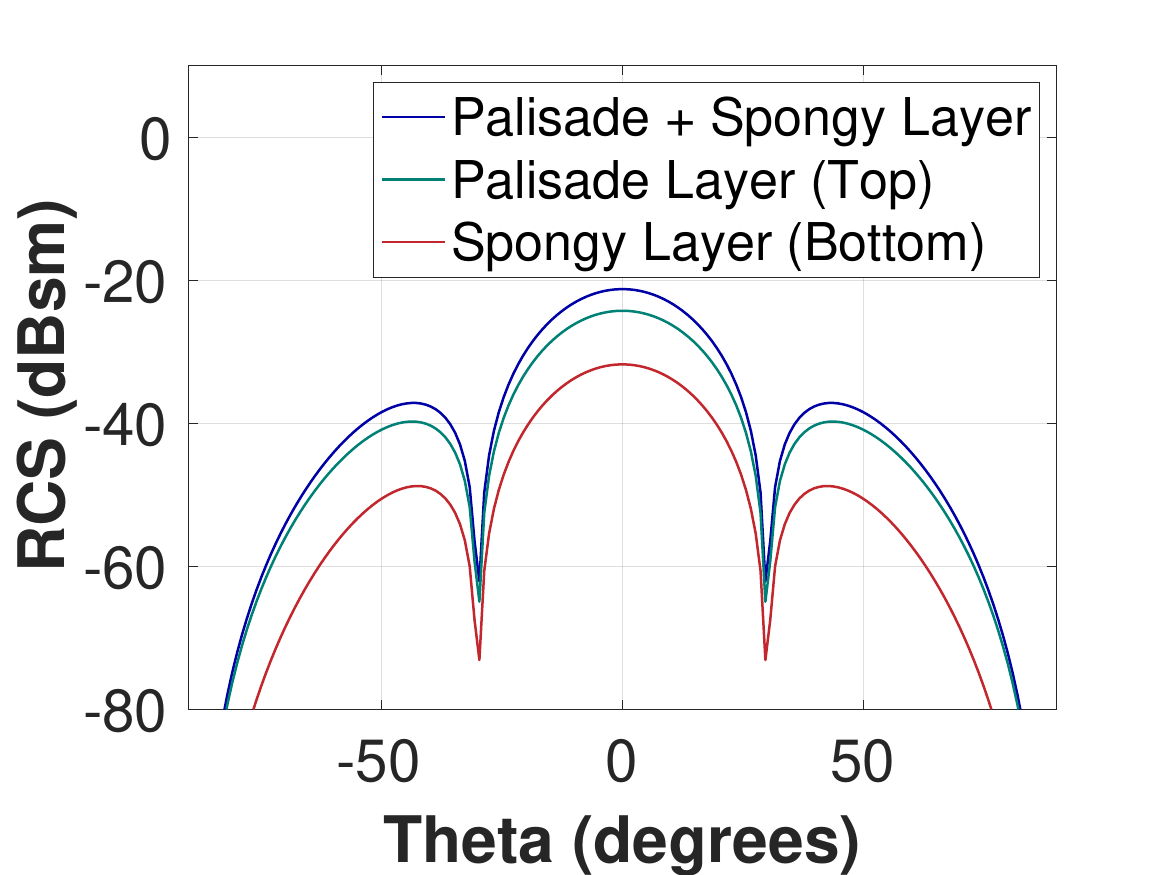}}
  \hspace*{0.5em}
  \subfloat[Reflections from each leaf layer at low RWC\label{fig:methodrcslayersrwc50}]{\includegraphics[width=0.48\linewidth]{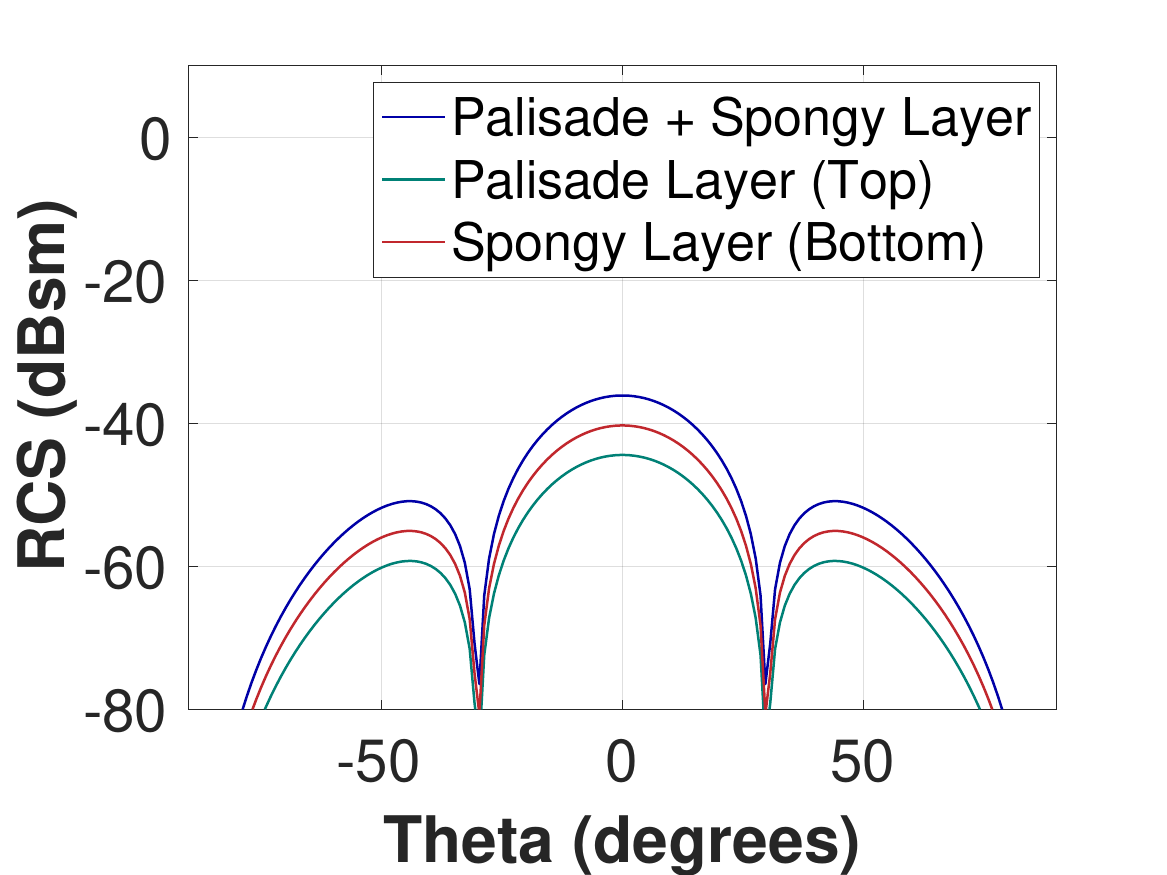}}
  \caption{Impact from different layers in the leaf on the radar reflection at high/low RWC}
  \label{fig:methodrcslayers}
  \Description{This graph shows the backscattered RCS for a rectangle leaf structure with two layers. The RCS is computed for vertical and horizontal polarization based on Equation 14 from \cite{sarabandi_millimeter_1990}.}
\end{figure}

For high RWC, we use the permittivity values of the palisade and spongy layers given in~\cite{sarabandi_millimeter_1990}, which assumes that the palisade layer has four times more water than the spongy layer. For low RWC, we assume that moisture level of the palisade layer has reduced to that of the spongy layer; in this case, the leaf essentially consists of one layer where the permittivity of both layers is the permittivity of the spongy layer. The results are shown in \ref{fig:methodrcslayersrwc100} and \ref{fig:methodrcslayersrwc50} for high and low  RWC. Each plot shows the contributions of surface and volumetric scattering to the RCS. We can see that for high RWC, the scattering from the palisade layer dominates while the scattering from the spongy layer dominates for lower RWC. Moreover, both figures show that the RCS depends on the angle of incidence, or in the context of our work, beamforming direction. 

Compared to prior work~\cite{hoog_60_2022}, \sn factors the surface and volumetric scattering effects resulting from the rough surface and internal structure of the leaf to increase the accuracy and robustness of leaf WC estimation methods based on the RSS ($A_d$) and AoA measurements from the radio waves with different incident angles.

\subsection{Receive Beamforming}
Eq.~(\ref{eqn:snells_law}) in Section~\ref{itm:background_leaf_structure} above shows that the RSS is affected by the AoA of the signal. To improve the leaf WC estimation accuracy, \sn incorporates receiver beamforming into the model to factor in the angle of arrival when a radio wave travels from air to the leaf and from leaf back to the air.
Specifically, the signal processing pipeline of \sn includes the Capon beamforming algorithm, also known as Minimum Variance Distortionless Response (MVDR) beamforming, to estimate the AoA of Rx radio signal. Capon beamforming algorithm processes the received antenna signals through weighting and summation, to focus the echoes towards a particular desired direction and minimize power from other directions, and provides a better resolution in angle estimation than other methods~\cite{xu_simultaneous_2022}. 
Formally, the Capon beamforming is defined as: 
\begin{equation}\label{eqn:aoacapon}
  W_{cap} = \frac{R_{xx}^{-1} \; a(\xi)}{a^H(\xi) \; R_{xx}^{-1} \; a(\xi)},
\end{equation}
where $W_{cap}$ \nomenclature{$W_{cap}$}{Capon beamforming weights for antenna array} is the weight vector to be applied to the measurements from receiver antenna array, $R_{xx}$ \nomenclature{$R_{xx}$}{Capon beamforming covariance matrix} is the covariance matrix of the received signal, $a(\xi)$ \nomenclature{$a(\xi)$}{Capon beamforming steering vector} is the steering vector in the look direction, and $H$ \nomenclature{$H$}{Capon beamforming Hermitian transpose} is the Hermitian transpose.

\begin{figure}[t]
  \vspace{-1em}
  \centering
   \subfloat[][Signal power for each Rx angle ranging from $-20^{\circ}$ to $+20^{\circ}$ \label{fig:methodavocadoleafair17rxbeamforming}]{\includegraphics[width=0.49\linewidth]{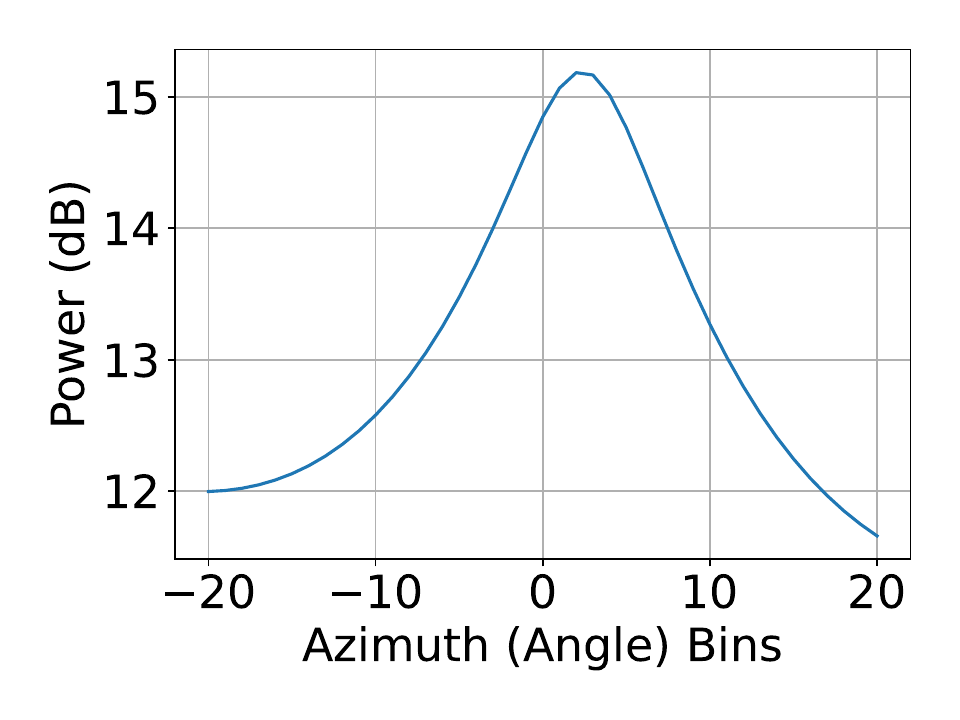}} 
   \hfill
   \subfloat[][Reflected power vs AoA\label{fig:resultsavocadoleafair14dbaoafiltered}]{\includegraphics[width=0.49\linewidth]{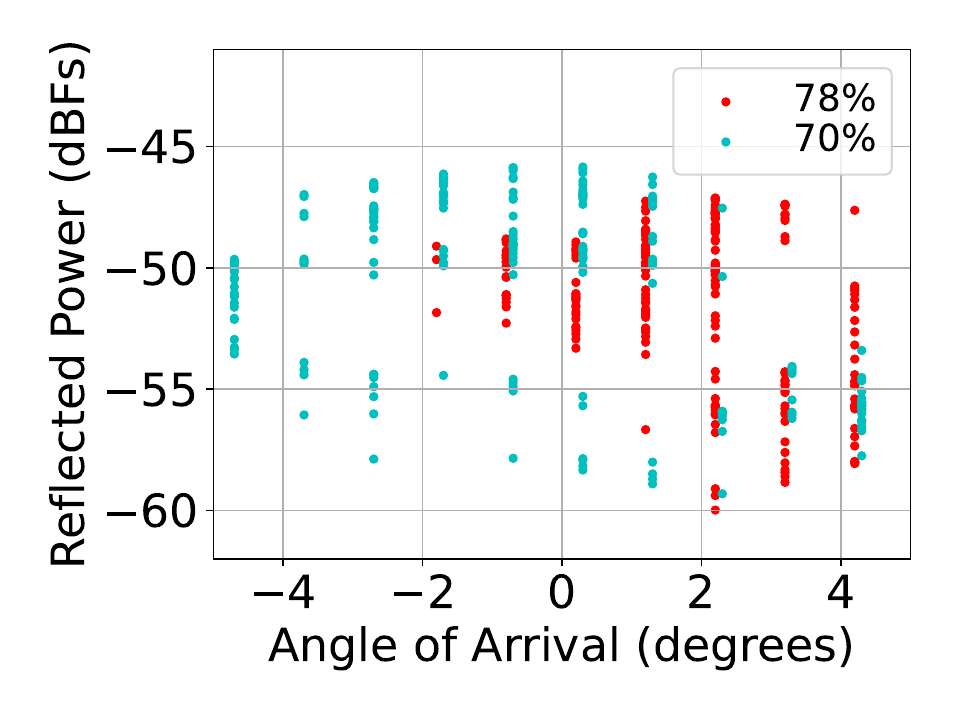}}
  \caption{The impact of AoA to Rx power strength.}
  \label{fig:AoAImpact}
\end{figure}

For a mmWave radar that has $\kappa$ \nomenclature{$\kappa$}{Number of receiver antennas in radar array} receiver antennas, the steering vector $a(\xi)$ is given by:
\begin{equation}\label{eqn:aoasteeringvector}
 a(\xi) = \begin{bmatrix}
                    1 \\
                    \mathrm{e} \, ^{j\frac{2\pi}{\lambda}(sin \xi) \, s} \\
                    \mathrm{e} \, ^{j\frac{2\pi}{\lambda}(sin \xi) \, 2s} \\
                    ...\\
                    \mathrm{e} \, ^{j\frac{2\pi}{\lambda}(sin \xi) \, (\kappa -1 )s}
              \end{bmatrix},
\end{equation}
where $\xi$ \nomenclature{$\xi$}{Capon beamforming desired angle} is the desired angle and $s$ is the space between
two neighboring antennas. The AoA of the Rx radio wave is obtained from the direction with the highest power from Eq.~(\ref{eqn:aoasteeringvector}), using a span of $\xi$ for the angle bins. Figure~\ref{fig:methodavocadoleafair17rxbeamforming} shows an example of the Rx beamforming output across a leaf surface, spanning from $-20^{\circ}$ to $+20^{\circ}$, where $2^{\circ}$ is the highest power angle bin and selected as the AoA of the Rx radio wave.

Figure \ref{fig:resultsavocadoleafair14dbaoafiltered} shows an example scatter plot of the signal strength against its correspondent AoA value of Rx radio waves for 
two different leaf RWC levels (i.e., 70\% and 78\%). The figure shows that the leaf with 78\% RWC level has peak RSS values around the AoA of 2º and 3º, while those of the 70\% RWC are around 0º and -1º. This makes the different RWC levels more differentiable than looking at RSS values only. Therefore, combining the signal strength and the AoA of Rx radio waves may assist in better estimating the leaf WC levels because it adds another dimension to the model.

Since it is challenging to measure the geometry of the surface of a leaf and its internal structure (see Figure~\ref{fig:methodleafstructure}), the precise relationship among the Tx beam steering angle, RSS and AoA is mathematically intractable in practice. Instead, we propose to use a DL network to model such linear and non-linear relationships, which will be introduced next.

\begin{table*}[t]
  \centering
  {
  \caption{Input and Output dimensions of the FCL in LM-Net.}
  \label{tab:dlmodelparameters}
  \begin{tabularx}{0.7\linewidth}{c|c|c|c|c|c|c|c|c}
    \toprule 
      \multicolumn{3}{c|}{\textbf{Location features}} & \multicolumn{3}{c|}{\textbf{RSS features}} &
      \multicolumn{3}{c}{\textbf{Adaptive Fusion}}\\
    \hline
      Layer & Input & Output & 
      Layer & Input & Output &
      Layer & Input & Output \\
    \midrule
      1 & $\iota \times 5$ & $\iota \times 16$ & 1 & $\iota \times \kappa \times 3$ & $\iota \times 64$ & 1 & $\iota \times 256$ & $\iota \times 32$\\
      2 & $\iota \times 16$ & $\iota \times 64$ & 2 & $\iota \times 64$ & $\iota \times 128$ & 2 & $\iota \times 32$ & $\iota \times 1$\\
      3 & $\iota \times 64$ & $\iota \times 128$ & 3 & $\iota \times 128$ & $\iota \times 256$ & 1 & $\iota \times 256$ & $\iota \times 32$\\
      4 & $\iota \times 128$ & $\iota \times 256$ & 4 & $\iota \times 256$ & $\iota \times 256$ & 2 & $\iota \times 32$ & $\iota \times 1$\\
      5 & $\iota \times 256$ & $\iota \times 256$ &   &     &     &   &     &   \\
     
       \bottomrule
  \end{tabularx}
  }
\end{table*}

\subsection{\sn Deep Learning Model}

The \sn network, called Leaf Moisture Network (LM-Net), is designed with a feature extractor and a decoder for regression (i.e., RWC value estimation). In addition to the AoA and RSS information from the Rx antennas, the feature extractor of LM-Net also attempts to extract the Tx steering angle information, since as previously mentioned, are informative to the leaf RWC. Finally, the decoder will estimate and output the RWC based on these features. 
Figure~\ref{fig:methodoverviewsystemworkflow} shows the architecture of LM-Net, which 
consists of four blocks: Location feature extractor, RSS feature extractor, Adaptive fusion, and the RWC predictor at the output.


\vspace{-0.2cm}
\subsubsection{Feature Extractor}\label{subsubsec:featureExtractor}

The inputs of LM-Net are: distance (or range bin) $d_t$; three AoA's  ($\theta_{d_{t-1}}$, $\theta_{d_t}$ and $\theta_{d_{t+1}}$) calculated at $d_t$ and its two neighboring range bins $d_{t-1}$ and $d_{t+1}$; Tx steering angle ($\eta$); and, RSS in the range bins $d_{t-1}$, $d_t$ and $d_{t+1}$ from all receiving antennas. In summary, the dimension of the inputs is $\iota \times 5$ ($\eta$, $\theta_{d_{t-1}}$, $\theta_{d_t}$, $\theta_{d_{t+1}}$ and ${d_{t}}$ ) for the Location input and $\iota \times \kappa \times 3$ for RSS input, where $\iota$ \nomenclature{$\iota$}{Number of transmit steering angles} is the number of Tx steering angles. For example, if we use $\iota = 11$ steering beams angles and $\kappa = 4$ Radar Rx antennas, the input dimension of LM-Net will be $ 11 \times 5 + 11 \times 4 \times 3 = 187$. Then, LM-Net will encode the relationship among the input into a latent representation (i.e., feature extraction).

Fully Connected Layer (FCL) blocks are often used for feature extraction purpose in many applications~\cite{qi_pointnet_2017, 10.1145/3458864.3467679}. Similarly, LM-Net makes usage of
FCL blocks as the feature extractor for leaf WC estimation and formulates it as a regression problem. Here, the feature extractor has two separated groups of FCL blocks: \textbf{Location features}, which include Tx steering angles and AoA, and \textbf{RSS features}, each of which represents different interaction between the radio signals and the leaf. This is because the radio wave Rx signal and AoA measurements from different Tx steering angles interact with ``rough'' leaf surface and its complex internal structure differently (see Section~\ref{itm:background_leaf_structure} above for the details). We must consider these factors in the feature extractor design.

Since the dimensions of Tx steering angles, Rx radio signal AoA and distance are $\iota$, $\iota \times 3$ and $\iota$ respectively, LM-Net concatenates them, i.e., \textbf{Location features}, to a dimension of $\iota \times 5$ before flattening the concatenated vector, which will be input into the neural network. Then, a series of FCL will extract the initial feature vector, before outputting the final feature vector. Similarly, for the \textbf{RSS features}, LM-Net concatenates the RSS from $\kappa$ Rx antennas for each steering angle and each of 3 range bins ($d_{t-1}$, $d_{t}$ and $d_{t+1}$), before flattening it to a vector with dimension of $\iota \times \kappa \times 3$ and inputting the vector to the neural network. The dimension of the output vectors of both \textbf{Location features} ($F_a$) \nomenclature{$F_a$}{LM-Net Location features output vector} and
\textbf{RSS features} ($F_r$) \nomenclature{$F_r$}{LM-Net RSS features output vector} after feature extraction is $\iota \times 256$. Table~\ref{tab:dlmodelparameters} summaries the detailed configuration of LM-Net FCL.

In LM-Net feature extractor, each layer performs dot product of input data with
a weight matrix $\mathcal{W}$ as:
\begin{equation}\label{eqn:dl_linear_block}
\mathcal{Z}^{[i]} = \mathcal{W}^{[i]}\times \mathcal{Z}^{[i-1]} + b^{[i]},
\end{equation}
\nomenclature{$\mathcal{W}^{[i]}$}{LM-Net weight matrix for the $i^{th}$ feature layer} \nomenclature{$\mathcal{Z}^{[i]}$}{LM-Net feature space vector for the $i^{th}$ layer} \nomenclature{$b^{[i]}$}{LM-Net bias vector for the $i^{th}$ feature extractor}  
where $\mathcal{W}\in \mathbb{R}^{l_i \times l_{i-1}}$, $l_i$ is the size of the $i$ layer output, and $b$ is the bias vectors. The input is the raw Tx steering angle, Rx signal AoA and signal strength values, in the first FCL, i.e.,  $i = 1$. The values of the elements in $\mathcal{W}$ and $b$ are randomly generated initially, and are learned during the training process automatically. This process is followed by a 1D Batch Normalization \cite{ioffe_batch_nodate} as:
\begin{equation}\label{eqn:dl_batchnorm}
\mathcal{Z'}^{[i]} = \frac{\mathcal{Z}^{[i]} - E[\mathcal{Z}^{[i]}]}{\sqrt{Var[\mathcal{Z}^{[i]}] }} \times \gamma + \beta,
\end{equation}
where $\gamma$ \nomenclature{$\gamma$}{LM-Net parameter in the feature space $\mathcal{Z}$} and $\beta$ \nomenclature{$\beta$}{LM-Net parameter in the feature space $\mathcal{Z}$} are parameters from the feature space $\mathcal{Z}$. Finally, the ReLu non linearity is applied to the $\mathcal{Z'}$ as:
\begin{equation}\label{eqn:dl_relu}
ReLU(\mathcal{Z'}) = max(0, \mathcal{Z'}^{[i]}),
\end{equation}
 which replaces the negative elements in the $\mathcal{Z'}$ 
 with $0$s.

\subsubsection{Adaptive Fusion}
The adaptive fusion in LM-Net automatically generates weights for the \textbf{Location features} and \textbf{RSS features} produced above to fuse them and obtain a robust WC output. Specifically, for both \textbf{Location features} and \textbf{RSS features}, LM-Net use a two layer FCL, i.e.,
Eq.~(\ref{eqn:dl_linear_block}), which is followed by an activation function ReLu, i.e., Eq.~(\ref{eqn:dl_relu}), with an input value dimension of $\iota \times 256$ and output
value dimension of $\iota \times 32$ (first layer) and $\iota \times 1$ (second layer), respectively.
The results of the FCL will be used as the weights of the \textbf{Location features} and \textbf{RSS features}
as:
\begin{equation}\label{eqn:dl_adaptive_weights}
  F_{m} = \omega_{a} \times F_{a} + \omega_r  \times F_{r},
\end{equation}
where $\omega_{a}$ \nomenclature{$\omega_{a}$}{LM-Net weights for the AF Location features} and $\omega_r$ \nomenclature{$\omega_{r}$}{LM-Net weights for the AF RSS features} are the weights for \textbf{Location features} and \textbf{RSS features} generated by the two-layer FCL respectively.

\subsubsection{Regression layer}
The regression layer is simply an FCL, i.e., Eq.~(\ref{eqn:dl_linear_block}),  accepting $F_m$ \nomenclature{$F_m$}{LM-Net adaptive fusion output vector} from the concatenated output from the Adaptive Fusion operation discussed above, and output/predicts \textbf{$\hat{RWC}$} \nomenclature{$\hat{RWC}$}{LM-Net output prediction value}. Therefore, the input dimension of the FCL is that of $F_m$, i.e., $\iota \times 256$, and the dimension of the output
of the FCL is one. Finally, we used Mean Squared Error (MSE) as the training loss ($\mathcal{L}$) as:
\begin{equation}\label{eqn:dl_loss_function}
  \mathcal{L} = ({RWC}_i-\hat{RWC}_i)^2.
\end{equation}

\subsubsection{Training Parameters}
An AdamW optimizer~\cite{loshchilov_decoupled_2019} is used to train the network with the learning rate automatically selected, using a PyTorch Lightning~\cite{noauthor_pytorch_nodate} module, but defaults to 0.005 with the weight decay set to $10^{-5}$. A learning rate scheduler is used to decay the learning rate by a multiplicative factor of 0.8 every 2 epochs. To stop the model over-fitting, an early stopping function is used to terminate the training if the testing loss does not continually decrease. To ensure reproducible results between runs, a seed value was set for the model dataloader worker. The batch size is set to 256 and the network is trained for 80 epochs.


\begin{table}
\centering
  \caption{TI AWR1843 Radar Chirp Parameters}
  \label{tab:radarchirpparameters}
  \begin{tabularx}{\linewidth}{l|l||l|l} 
  \toprule
    Parameters & Specs. & Parameters & Specs. \\ 
    \hline
    Frequency (GHz) & 77-81 & Slope ($MHz/\mu s$)  & 18.32 \\
    Idle time ($\mu s$) & 7 & Sample rate (ksps) & 5,000 \\
    Ramp start ($\mu s$) & 7 & No. chirps & 32 \\
    Ramp end ($\mu s$) & 212.8 & Chirp time ($\mu s$) & 10 \\
    No. ADC samples & 1,024 & Frame length (ms) & 350 \\
    Bandwidth (GHz) & 3.75 &  &  \\
  \bottomrule
\end{tabularx}
\end{table}

\section{Evaluation}\label{sec:evaluation}

\subsection{Goals, Methodology and Metrics}\label{subsec:goalsMethodologyMetrics}

\subsubsection{Goals}
Our goals are three-folds: first, the performance of \sn in RWC estimation
in both controlled detached leaves and live plants (indoors and outdoors); second, bench-marking \sn against the SOTA approach~\cite{hoog_60_2022}; and third, understanding the contributions of different components of \sn to its performance. 

\subsubsection{Methodology}

\paragraph{\textbf{Prototype}}
To evaluate the performance of \sn, we implement a prototype using a COTS AWR1843Boost radar and a DCA1000EVM data collection board, both from Texas Instruments (TI). The AWR1843 is a 77 GHz mmWave radar with $m = 3$ Tx (Tx1, Tx2 and Tx3) and $\kappa = 4$ Rx (Rx1, Rx2, Rx3 and Rx4) with programmable phase rotators in the transmit path to enable beamforming \cite{noauthor_awr1843boost_nodate}. The DCA1000 provides real-time data capture from the AWR1843 to a computer, which saves the raw ADC values into a binary file \cite{noauthor_dca1000evm_nodate}. Table \ref{tab:radarchirpparameters} shows the FMCW chirp parameters used in the \sn prototype.

\nomenclature{$m$}{Number of transmit antennas in the radar array}

\paragraph{\textbf{Tx beam steering phase calculation}}
Eq.~(\ref{eqn:aoaphaseomega}) earlier shows that the Tx Beam Steering angle, which is a
part of LM-Net \textbf{Location features} as discussed in Section~\ref{subsubsec:featureExtractor}, 
depends on the inter-antenna spacing $s_m$ and 
can be controlled the phase offset ($\phi$) between the antennas. Since the spacing between Tx1 and Tx2 ($s_2$) in our prototype is $\lambda$ and the space between Tx1 and Tx3 is ($s_3$) $2 \times \lambda$, we may control the Tx beam Steering angle ($\eta$) of our \sn prototype via:
\begin{equation}\label{eqn:mmwavephaseoffsetsawr1843}
  \overset\rightharpoonup{\phi} = [\phi_1 \; \phi_2 \; \phi_3]  = \left[ 0 \quad 2\pi\sin\eta \quad  4\pi\sin\eta \right].
\end{equation}
We used up to eleven, i.e., max$(\iota) = 11$, Tx steering angles in our evaluation, starting from $-10^{\circ}$ to $+10^{\circ}$ with increments of $2^{\circ}$.

\begin{figure}[t]
  \centering
  \subfloat[Avocado]{\includegraphics[width=0.35 \linewidth]{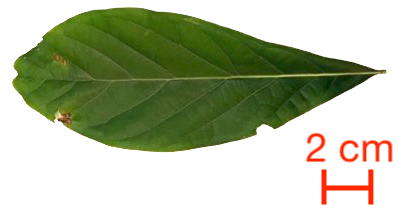}}
  \hspace*{1.0em}
  \subfloat[Rubra]{\includegraphics[width=0.15\linewidth]{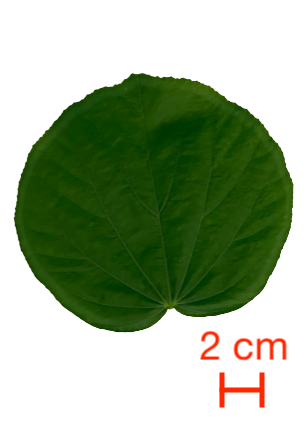}}
  \hspace*{1.0em}
  \subfloat[Bull bay\label{fig:leafTypesBullBay}]{\includegraphics[width=0.45\linewidth]{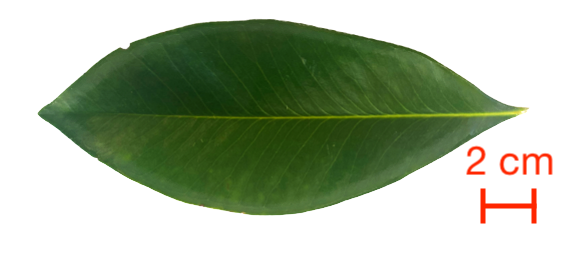}}
  \caption{Leaf types used in experiments.}
  \label{fig:leafTypes}
\end{figure}

\paragraph{\textbf{Leaf types and WC ground truth}}
We test three types of leaves as shown in Figure~\ref{fig:leafTypes}: Persea americana (Avocado),
Hibiscus tiliaceus rubra (Rubra), and Magnolia grandiflora (Bull bay), which cover different sizes, shapes, and surface smoothness.

\begin{figure}[t]
  \centering
  \vspace{-1.0\baselineskip}
  \subfloat[100\%]{\includegraphics[angle=0,width=0.17\linewidth]{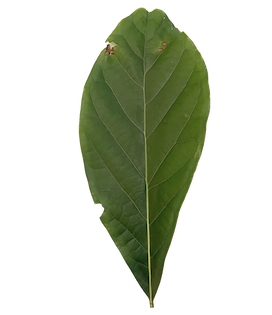}}
  \hfill
  \subfloat[90\%]{\includegraphics[angle=0,width=0.16\linewidth]{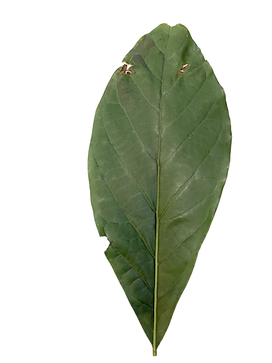}}
  \hfill
  \subfloat[80\%]{\includegraphics[angle=0,width=0.17\linewidth]{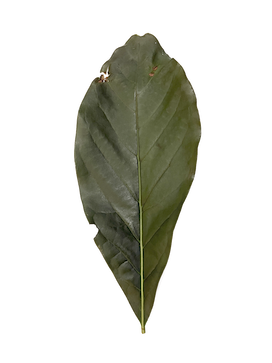}}
  \hfill
  \subfloat[70\%]{\includegraphics[angle=0,width=0.16\linewidth]{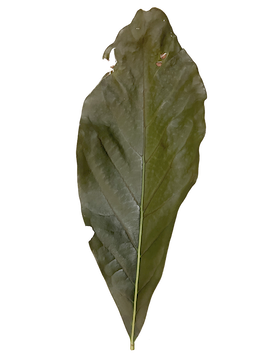}}
  \hfill
  \subfloat[60\%]{\includegraphics[angle=0,width=0.16\linewidth]{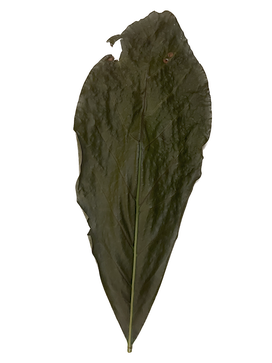}}
  \hfill
  \subfloat[50\%]{\includegraphics[angle=0,width=0.16\linewidth]{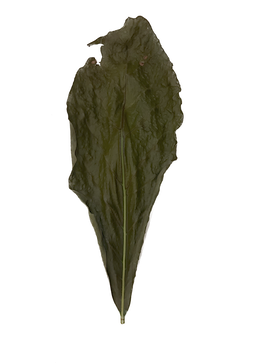}}
  \vspace{-1em}
  \caption{An example Avocado leaf at different RWC levels.}
  \label{fig:resultsavocadoleaf7dryingset}
  \Description{Shows the physical changes of the leaf at RWC 100\% 90\% 80\% 70\% 60\% 50\%}
\vspace{-0.5cm}
\end{figure}

 A common ground truth for accurately determining leaf WC is by measuring the weight from a detached leaf \cite{peden_model_2021, dey_paper_2020, afsharinejad_variability_2017, ulaby_microwave_1987}, and the RWC may be calculated as:
\begin{equation}\label{eqn:rwcmodified}
  RWC_m = \frac{f_w}{t_w} \times 100,
\end{equation}
where $f_w$ is the current weight of the leaf, and $t_w$ is the turgid weight (i.e., maximum leaf weight).
Figure~\ref{fig:resultsavocadoleaf7dryingset} shows an example Avocado leaf at different RWC levels.

\paragraph{\textbf{In-lab data collection}} We collect a total of 2,160 RWC samples of mmWave radio measurements of three different types of leaves (i.e., Avocado, Rubra and Bull bay) with the RWCs between 50\% and 100\%. 
Similar sized leaves were cut from the plant, the weight was measured to the nearest 0.01g and used as the starting weight $t_w$. The leaves were placed on top a styrofoam box during radar measurements to minimize background reflections.
\revisedtext{To increase the location and orientation robustness, the detached leaf was slightly-moved 20 times (for each RWC\%) to increase the spatial variability during training. The detached leaf was initially placed at boresight and adjusted within 5 degrees of boresight in both the azimuth and elevation plane, well within the radar field-of-view (FOV).}
Similar to the method in~\cite{hoog_60_2022}, the leaves were oven dried at $60^{\circ}C$ with radar measurements taken at 10\% RWC intervals, as shown in Figure \ref{fig:resultsavocadoleaf7dryingset}. The leaves were weighed after each round of drying to obtain $f_w$ for each RWC level. 

\paragraph{\textbf{In-lab Experimental settings}} 
The detached leaf was placed at distances 0.4m, 0.6m and 0.8m away from the radar with a leaf aspect angle of $0^{\circ}$. The distances were selected using the radar's $15^{\circ}$ azimuth beamwidth and the desired FOV being slightly larger than the leaves tested, i.e. between 10cm and 20cm. The majority of results in the paper use a distance of 0.6m which should be assumed as the default distance unless specified otherwise. At a distance of 0.6m, the radar FOV is approx 16cm wide ($\frac{0.08}{0.6} \approx \tan\left( \frac{15^{\circ}}{2} \right)$) which is slightly larger than all leaves tested. The leaf was placed within $\pm2cm$ of the radars boresight so that each radar measurement, which steered the Tx beam from $-10^{\circ}$ to $+10^{\circ}$ in increments of $2^{\circ}$, ensured the entire leaf was measured.  The \revisedtext{data collection procedure for each leaf was} conducted over a short period of time, approx. 1 hour, where the environmental conditions were relatively stable. It was assumed only one leaf was in the radar FOV and it was non-moving during the experiment.

\subsubsection{Metrics}
We use 10-fold cross-validation for the in-lab evaluation with the exception of the leaf distance experiment which used leave-one-group-out. We use MAE to evaluate the performance of \sn, which shows the difference between the estimated RWC and the ground truth.


\subsection{The Impact of Tx Beam Steering}
\begin{figure}[t]
    \centering
        \includegraphics[width=\linewidth]{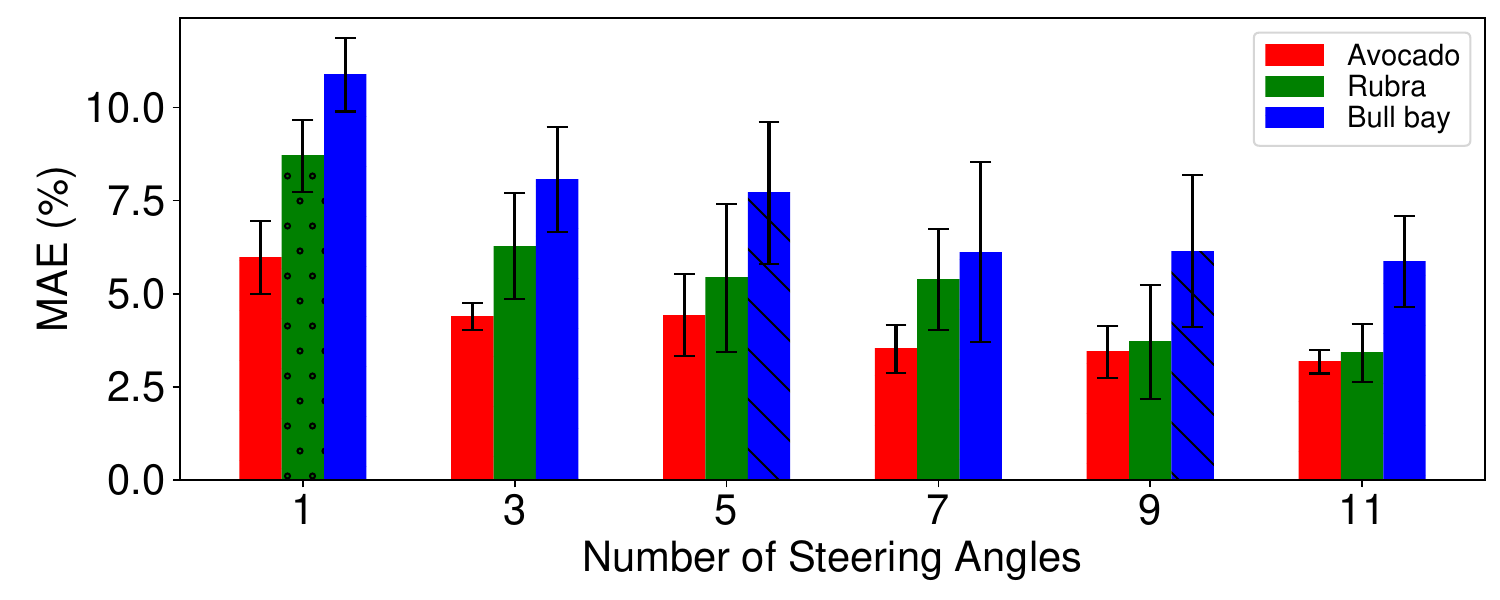}
        \vspace{-2em}
        \caption{Performance of \sn vs. different numbers of Tx steering angles.}
        \label{fig:results10foldsteeringanglesrmse}
        \vspace{-1em}
\end{figure}

Given that radio waves' interactions with leaves are influenced by their incident angle, as previously discussed in Section~\ref{itm:background_leaf_structure}, \sn steers the Tx beam across the leaf surface to obtain a more fine-grained leaf RWC estimation than prior work, which used a single angle only. Figure \ref{fig:results10foldsteeringanglesrmse} shows the performance of \sn improves for all leaf types by increasing the number of Tx beam steering angles. For example in the Rubra leaf, the MAE of RWC estimation reduces by more than 60\%, i.e., from approximately 8.7\% to approximately 3.41\%, when the number of Tx beam steering angles increases from one to eleven. Although there is a flat trend of error when using 5 and 7 Tx beam steering angles for Avocado and Rubra leaf respectively, the general trend shows the prediction accuracy improving with higher numbers of Tx beam steering angles. We note the RWC estimation performance improvement diminishes after the number of steering angles is more than ten, so will use eleven steering angles in our evaluation for the rest of the paper.


\subsection{Overall Performance}
\begin{figure}[t]
    \centering
        \includegraphics[width=1.01\linewidth]{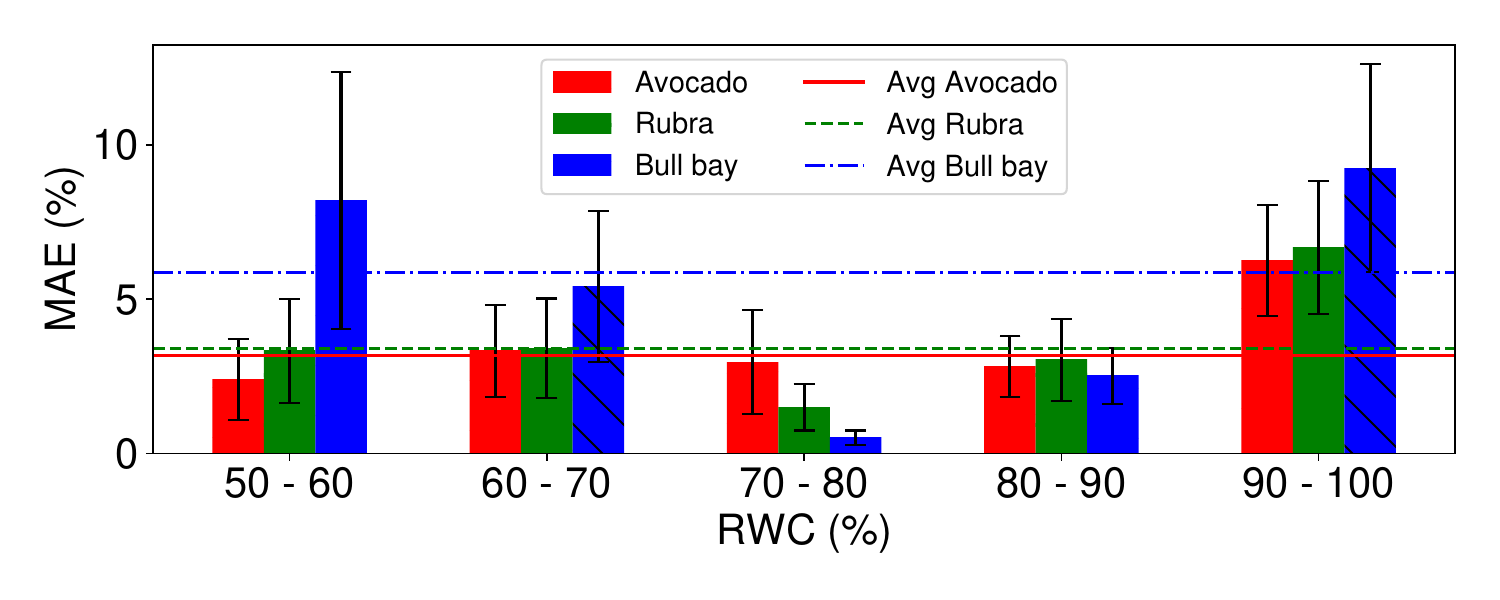}
        \vspace{-2em}
        \caption{The overall RWC estimation performance (MAE \%).}
        \label{fig:results10foldprediction}
        \vspace{-0.5cm}
\end{figure}

\begin{figure*}[t]
  \centering
  \vspace{-1em}
  \subfloat[Avocado]{\includegraphics[width=0.33\linewidth]{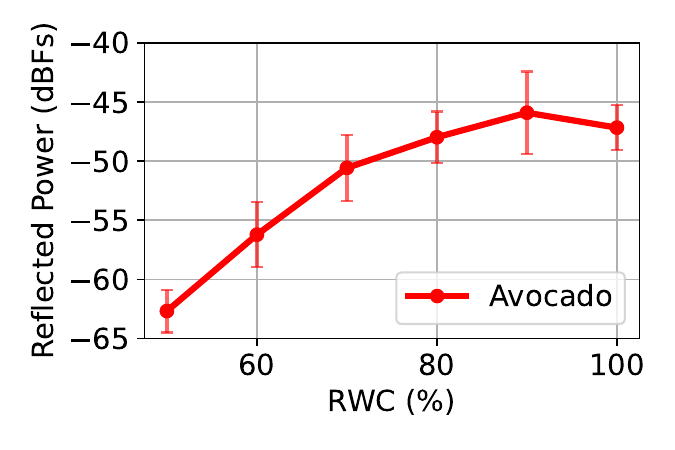}}
  \subfloat[Rubra]{\includegraphics[width=0.33\linewidth]{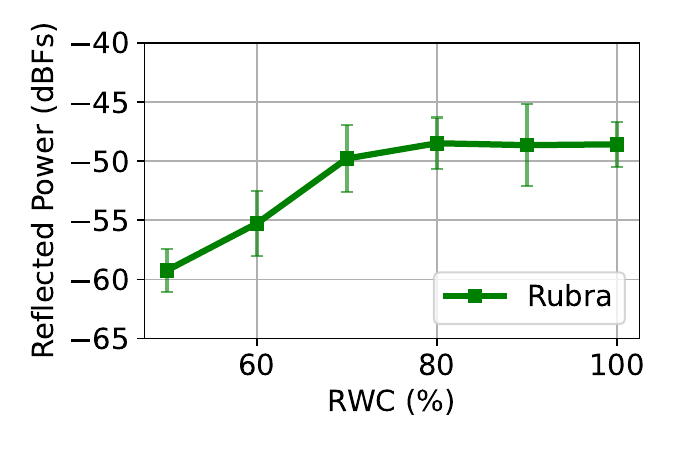}}
  \subfloat[Bull bay\label{fig:resultsrwcseriesBullBay}]{\includegraphics[width=0.33\linewidth]{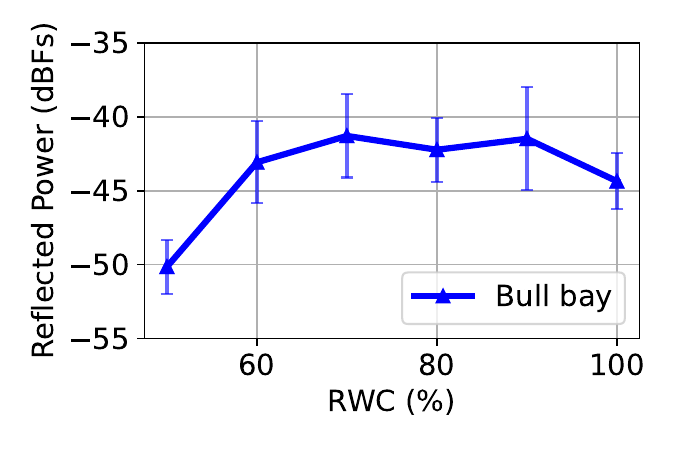}}
  \vspace{-1em}
  \caption{Reflected power vs RWC among three types of leaves.}
  \vspace{-1em}
  \label{fig:resultsrwcseries}
  \Description{This shows the mean reflected power for each RWC weight class for the three types of leaves.}
\end{figure*}

Figure~\ref{fig:results10foldprediction} shows the MAE results of \sn at different RWC levels for different types of leaves, grouped by RWC intervals of 10\%. It shows that, at all RWC levels and different types of leaves, the estimated RWC does not deviate too much from the ground truth. \sn performs better in the RWC estimation of the Avocado leaf (with an MAE of 3.17\%) and Rubra leaf (with an MAE of 3.41\%) than the Bull bay leaf (with an MAE of 5.87\%). This is because of the ``smoothness'' difference of the leaves (see Figure~\ref{fig:leafTypes}). Figure~\ref{fig:resultsrwcseries} further illustrates this behavior by plotting the average RSS against the RWC. For the Avocado and Rubra leaf, the RSS increases with the RWC, as surface scattering is a more dominant factor than volumetric scattering for radio signal reflections in ``smooth'' surfaces. On the other hand, the RSS vs. RWC levels of the Bull bay leaves are comparatively flat since volumetric scattering is a more dominant factor for radio signal reflections in ``rough'' surfaces. This indicates the importance of LM-Net model's \textbf{Location features}, in addition to the \textbf{RSS features} used in the SOTA approach~\cite{hoog_60_2022}, which will be investigated next.

\begin{figure}[t]
    \centering
    \captionsetup[subfloat]{farskip=3pt,captionskip=1pt}
        \centering
        \subfloat[LM-Net module performance.]{
        \includegraphics[trim={0.5cm 0cm 0.2cm 0.6cm},clip,width=0.5\linewidth]{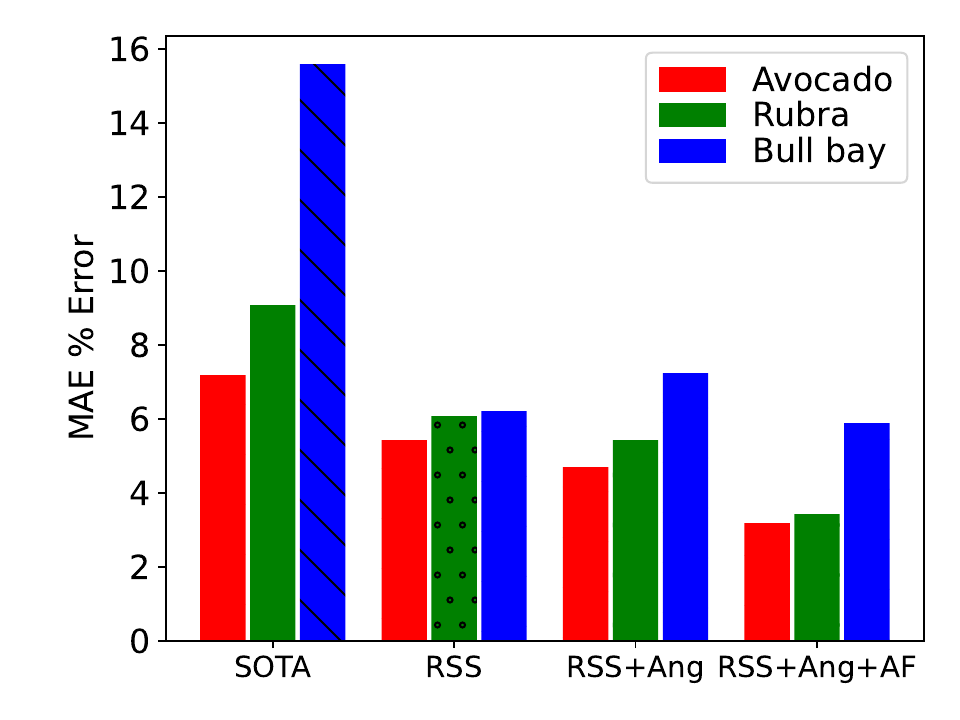}
        \label{fig:resultsdeeplearningablationmodules}
        }
        \centering
        \subfloat[LM-Net distance performance.]{
        \includegraphics[trim={0.49cm 0cm 0.2cm 0.8cm},clip,width=0.52\linewidth]{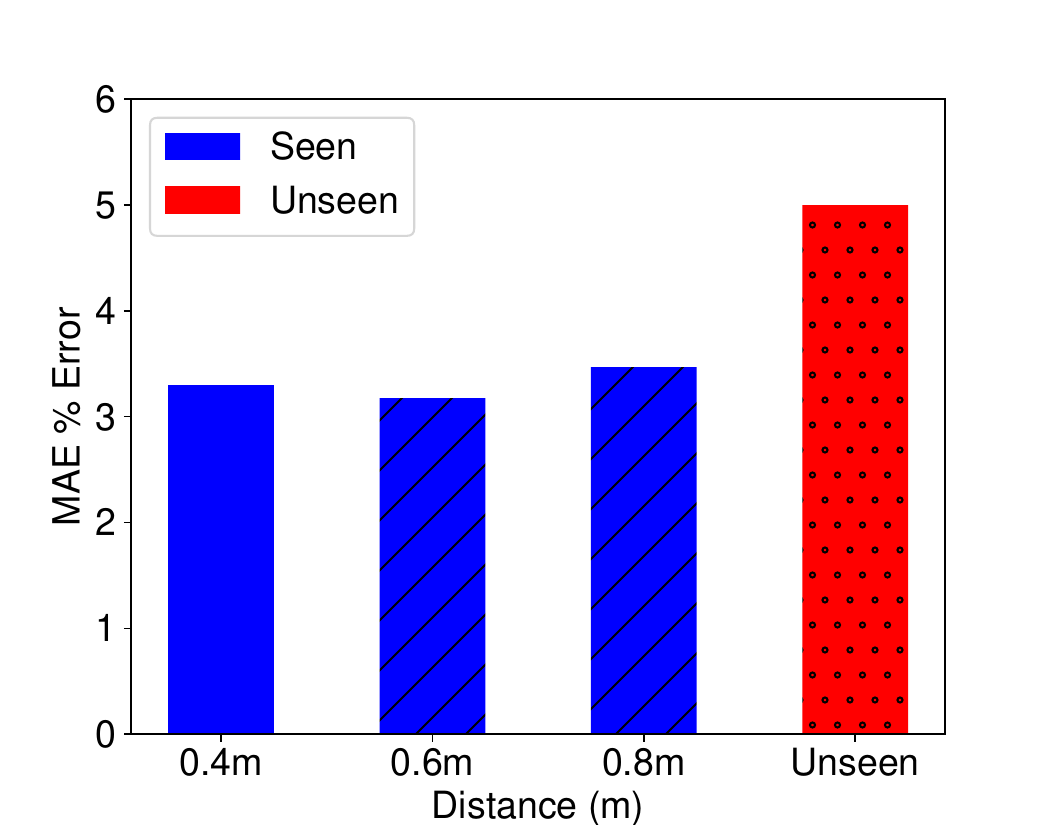}
        \label{fig:resultsdistancetestmae}
        }
    \caption{LM-Net Performance}
   \vspace{-2em}
\end{figure}

\subsection{LM-Net Performance Comparison with SOTA}
We investigate the impact of different LM-Net components
(see Figure~\ref{fig:methodoverviewsystemworkflow}), i.e., Location feature extractor (Ang), RSS feature extractor (RSS), and Adaptive Fusion (AF), to the performance of \sn in leaf RWC estimation. We consider the performance of the model with RSS features and the radio signal from a single
Tx steering angle at $0^{\circ}$ as the SOTA approach~\cite{hoog_60_2022}, which is based on the RSS from a single radio wave only. 
Figure~\ref{fig:resultsdeeplearningablationmodules} shows that 
the performance of the LM-Net model significantly improves with the addition of the Ang and AF components, compared to with RSS features only, i.e.,
SOTA approach~\cite{hoog_60_2022}. Including the \textbf{Location features} component highlights the informative features which are contained in the AoA information and can be used to improve the model's accuracy, e.g., the Avocado leaf MAE improved by 16.5\%, going from 6.48\% to 5.41\%. Adding the AF component, which dynamically fuses the RSS and the AoA information, further enhances the discriminatory ability of the network and provides the best performance of the model, e.g., the MAE of the Rubra leaf reduced by approximately 62.4\%, going from 9.07\% to 3.41\%. We note that for more challenging leaves with ``rough'' surfaces such as the Bull bay leaf (see Figure~\ref{fig:leafTypesBullBay} and~\ref{fig:resultsrwcseriesBullBay}), simply combining RSS and Location features
actually increases the RWC estimation errors, i.e., from 6.20\% to 7.23\%, \revisedtext{as the weights of the Location feature module have not been adjusted for the increased AoA sensitivity}; nevertheless, the LM-Net model in \sn (i.e., RSS + Ang + AF in the figure) produced the best performance of 5.87\% for the Bull bay leaf, which demonstrates the effectiveness of the adaptive fusion component in the LM-Net.


\subsection{Impact of Distance}
The performance of LM-Net for three different distances, i.e., 0.4m, 0.6m, and 0.8m, was evaluated. The MAE \% Error of \sn was consistent across all leaf distances when the distance was known to the model prior, as shown in Figure~\ref{fig:resultsdistancetestmae}. For the untrained, i.e. \textit{unseen}, leaf position, there was a drop of performance, with the MAE Error increasing from 3.17\% to 4.99\%. It should be noted that the performance of LM-Net using \textit{unseen} leaf positions is still better than the SOTA~\cite{hoog_60_2022} approach, shown in Figure~\ref{fig:resultsdeeplearningablationmodules}, which uses a fixed known distance.

\begin{figure}[t]
    \centering
    \captionsetup[subfloat]{farskip=4pt,captionskip=1pt}
    \subfloat[Predicted RWC vs Measured Soil VWC (indoor) \label{fig:resultsleaftypeliveindoortargetscalerpowertransformer}]{\includegraphics[trim={0.5cm 0.6cm 0.0cm 0cm},clip,width=0.46\linewidth]{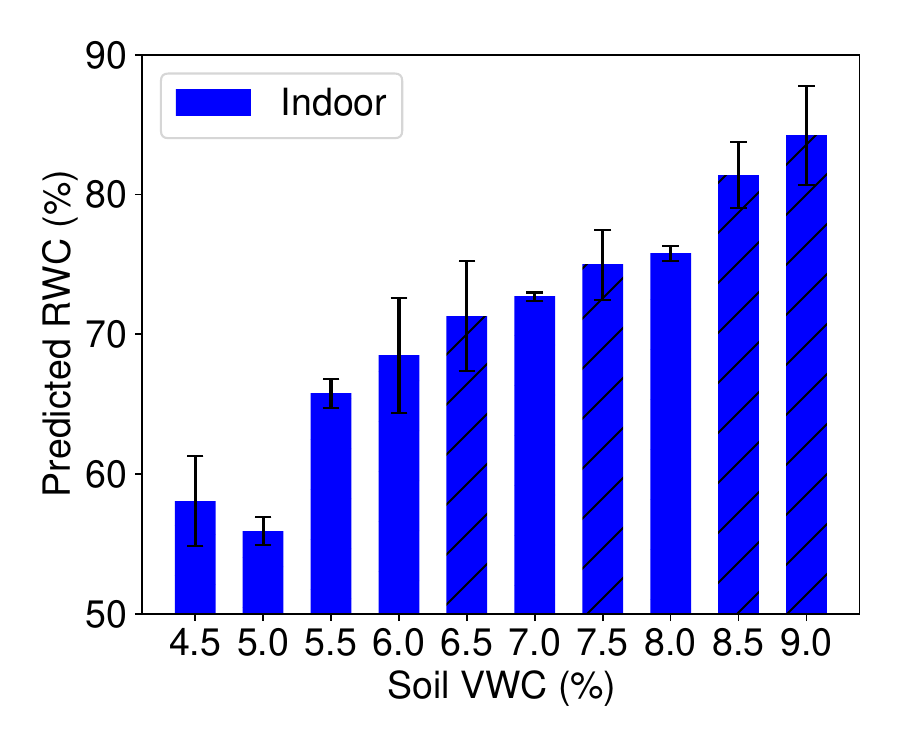}}
    \hspace*{0.5em}
    \subfloat[\revisedtext{Glasshouse farm experimental setup} \label{fig:resultsleaftypeliveplantexperimentsetup}]{\includegraphics[trim={0.5cm 1.0cm 0.0cm 0.0cm},clip,width=.46\linewidth]{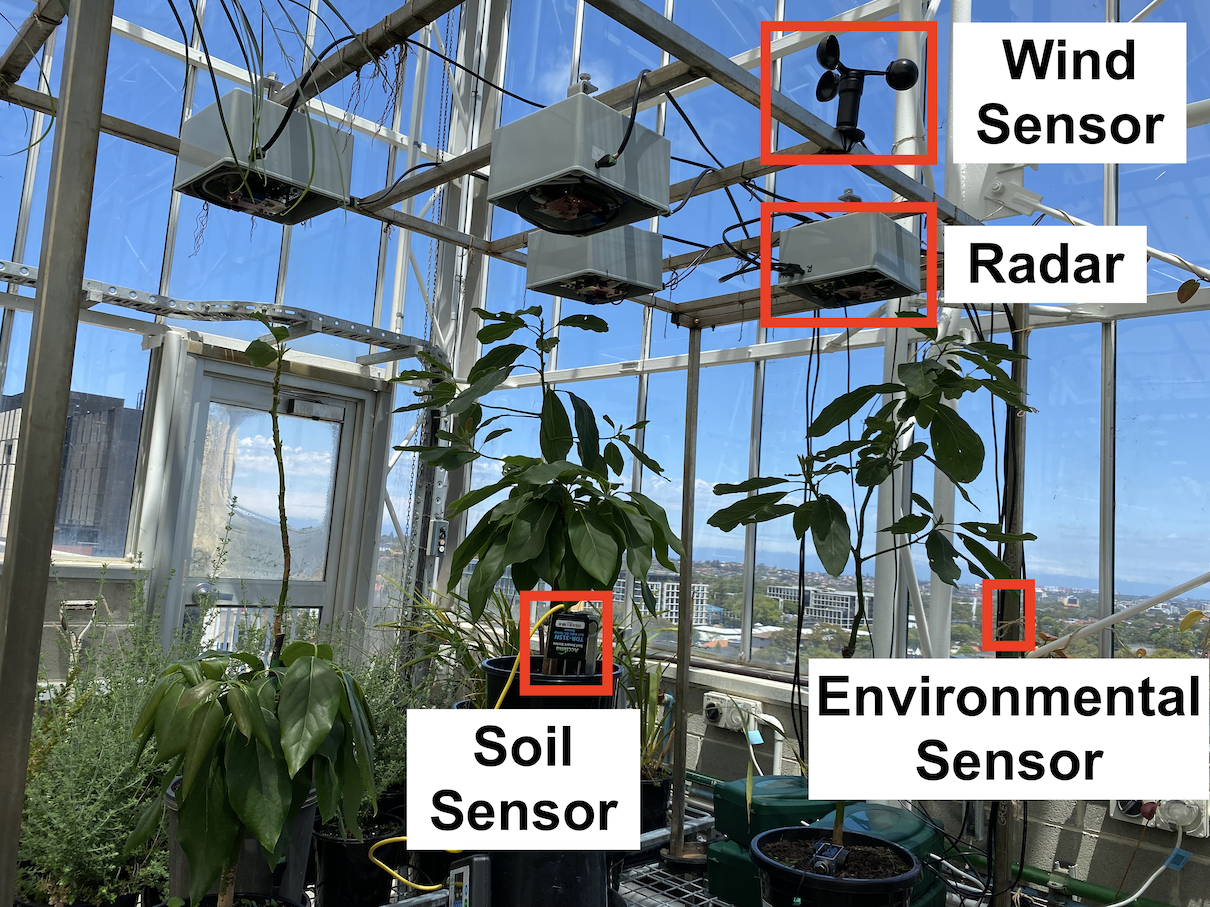}}
    \vspace{-1em}
    \caption{}
    \vspace{-2em}
    \label{fig:resultsleaftypelivetargetscalerpowertransformer}
\end{figure}

\subsection{Live Plant Experiment} 
For the benefit of usability and to validate the model's generability, the LM-Net model \textbf{was trained using the dataset from the in-lab detached leaves and tested on multiple live Avocado plants}. 
Measuring the RWC of an attached leaf is challenging and most measurement techniques use a proxy to estimate the leaf RWC value~\cite{khait_sounds_2023}. Following the method in~\cite{khait_sounds_2023}, we used soil moisture\footnote{Measured by Acclima SDI-12 https://acclima.com/sdi-12-sensor-reader-kit-user-manual/} as a proxy to estimate leaf WC in the live plant experiments since there is a strong relationship between soil moisture and leaf water potential~\cite{gaudin_linking_2017}. \revisedtext{Nevertheless, leaf-based approaches offer a more reliable water status estimation than soil-based approaches by removing the complex calculations required between the soil-plant-atmosphere interfaces \cite{AFZAL2017148}}.

\subsubsection{Indoor Environment}
An indoor experiment was set up using two Avocado plants
in our lab, where the plants were initially watered and then measured for 7 days without additional hydration. The temperature of the room was kept at $22^{\circ}C \pm 3^{\circ}C$ and the ViparSpectra XS1500 LED provided $500\mu mol/m^2/s$ of light, measured on the leaf~\cite{doi:10.1098/rsos.160592}, for photosynthesis between 6am and 8pm daily. The radar was positioned so its FOV would primarily capture a single leaf, with measurements taken every hour. Figure~\ref{fig:resultsleaftypeliveindoortargetscalerpowertransformer} shows there is a strong relationship between the predicted leaf RWC and the measured soil VWC, despite the LM-Net model was pre-trained to ensure its generability and usability.

\subsubsection{Glasshouse Experimental Farm Environment}
\revisedtext{An outdoor experiment using natural sunlight was set up in a glasshouse (see Fig.~\ref{fig:introleaftypeliveplantexperimentsetup}), where four Avocado plants were initially watered and then measured for 10 days without additional hydration by four radars shown in Figure~\ref{fig:resultsleaftypeliveplantexperimentsetup}. A radar was positioned so its FOV would primarily capture a single leaf, with measurements taken every hour.}

\textbf{Impact of environmental factors (temperature, humidity, and wind.)}
The temperature, humidity and wind speed were recorded in the \revisedtext{glasshouse} to study their impact on the performance of LM-Net model, using COTS IoT sensors~\cite{noauthor_nano_nodate, noauthor_wind_speed_sensor_voltage_type_0-5v__sku_sen0170-dfrobot_nodate} (see Figure~\ref{fig:resultsleaftypeliveplantexperimentsetup}), every 15 seconds. 

Our results show that temperature ($18^{\circ}C \pm 5.4^{\circ}C$) and humidity ($69\% \pm 13\%$) exhibited limited influence on the LM-Net prediction output; \revisedtext{however, extreme fluctuations were not tested. Monitoring a well-watered plant over several days showed that the prediction output stayed relatively constant. Only wind speed impacted the prediction output, as LM-Net was trained on the dataset collected from non-moving plant leaves. Although the glasshouse was mostly protected from wind, air conditioning and fans did occasionally introduce leaf flutter.} To limit the impact from higher wind speeds, radar took measurements when the wind speed is less than 0.1m/s only. 
\revisedtext{Figure~\ref{fig:resultsleaftypeliveoutdoortargetscalerpowertransformer} shows there is a strong relationship between the predicted leaf RWC and the measured soil VWC, similar to that in the indoor experiment.}

\begin{figure}[t]
  \captionsetup[subfloat]{farskip=2pt,captionskip=1pt}
  \centering
  \subfloat[RSS feature \label{fig:resultsablationextractosblocksmaea}]{\includegraphics[trim={0.6cm 0.5cm 0.5cm 0.5cm},clip,width=0.33\linewidth]{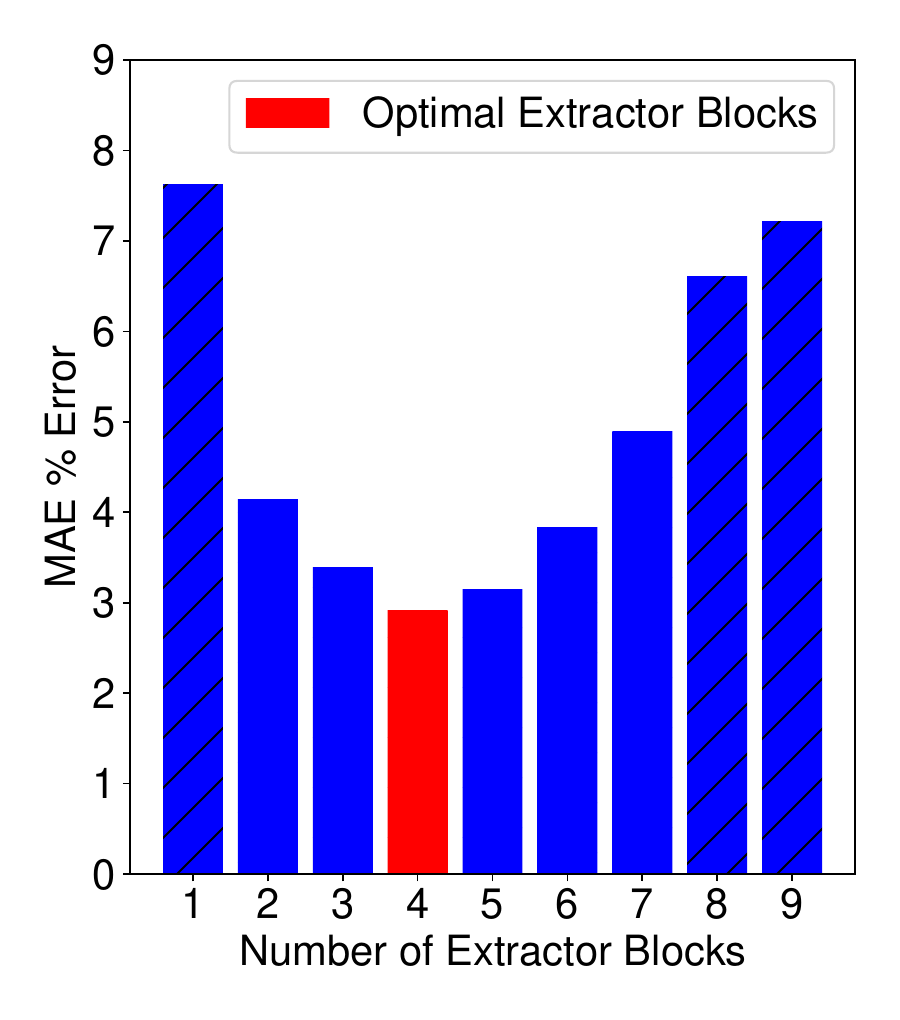}}
  \subfloat[Location feature \label{fig:resultsablationextractosblocksmaeb}]{\includegraphics[trim={0.6cm 0.5cm 0.5cm 0.5cm},clip,width=0.33\linewidth]{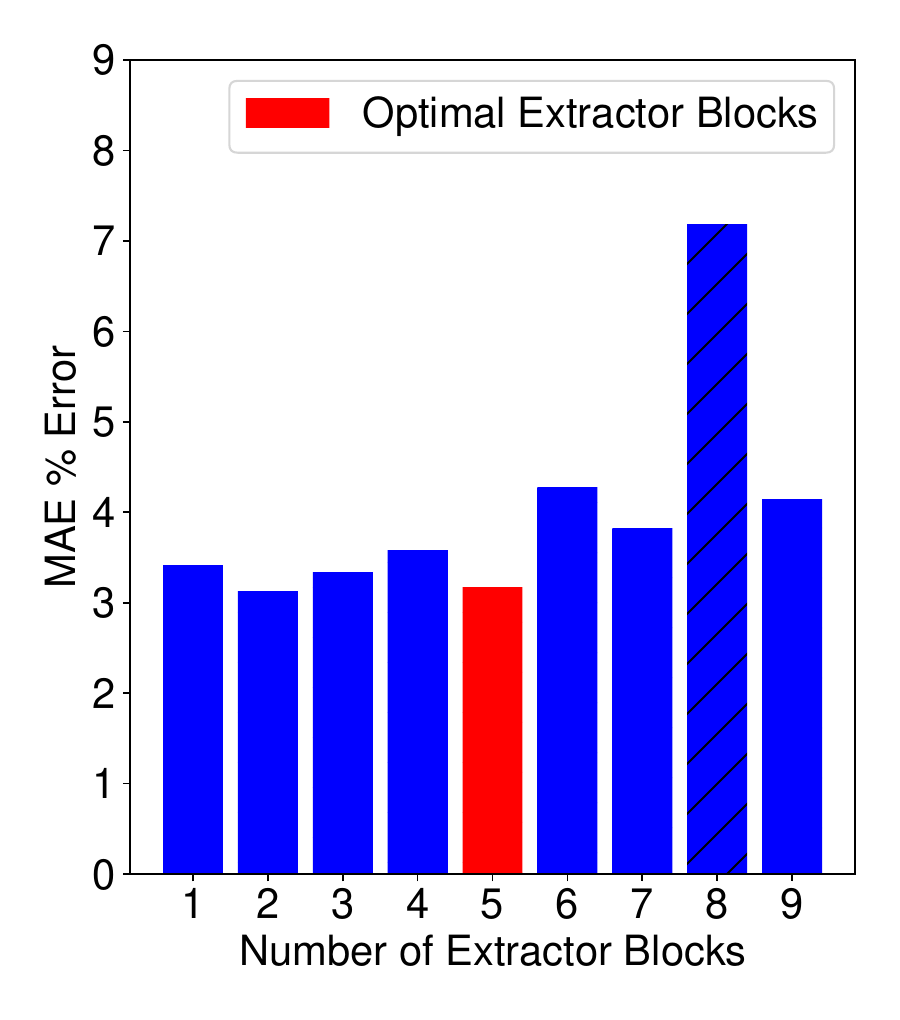}}
  \subfloat[Batch size \label{fig:resultsablationextractosblocksmaec}]{\includegraphics[trim={0.6cm 0.5cm 0.5cm 0.5cm},clip,width=0.33\linewidth]{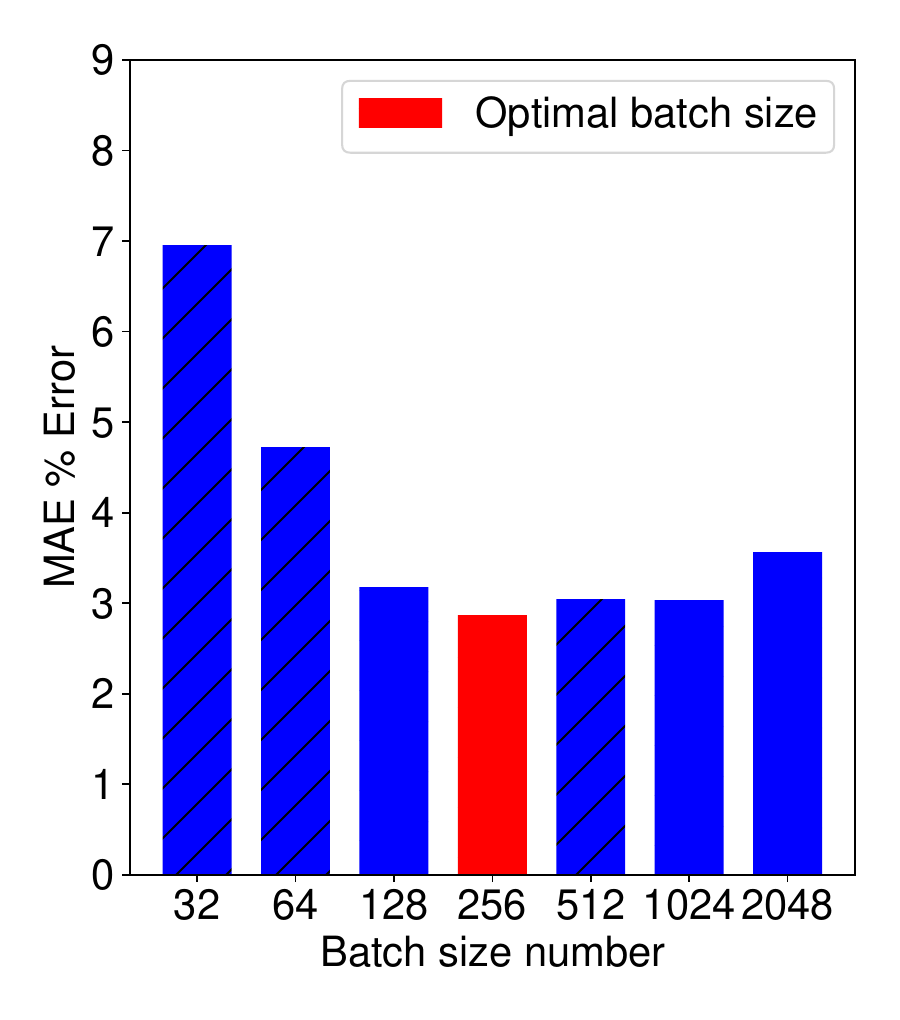}}
  \vspace{-1em}
  \caption{Ablation study: measuring the performance of different numbers of layers in the LM-Net Feature Extractor Blocks and different batch sizes.}
  \label{fig:resultsablationextractosblocksmae}
  \vspace{-1em}
  \Description{Graphs show the MAE error for different numbers of layers in the Feature Extractor Blocks where a) changed the RSS feature layers only and b) changed the Ang feature layers only.}
\end{figure}

\subsection{Ablation Study}
We study the impact of the different LM-Net model parameters to its performance. Figure~\ref{fig:resultsablationextractosblocksmae} shows its performance against different numbers of FCL blocks in the RSS and Location feature extractors, as well as different batch sizes in LM-Net for the Avocado leaves. The optimal performance (3.17\%) is achieved when the number of FCL blocks are 4 and 5 for RSS feature extractor and Location feature extractor respectively, and the batch size is 256, which is the configuration in the LM-Net model of our prototype implementation.

\subsection{Hardware Overhead}
We implement the LM-Net model in our prototype with the PyTorch Lightning framework~\cite{noauthor_pytorch_nodate}, which is a lightweight wrapper of PyTorch, and a popular framework for neural networks. The training of LM-Net is completed on a shared computational cluster running Rocky Linux 8.6, 16-core Intel Xeon CPU, a Tesla V100-SXM2 GPU and 46GB of RAM. The overall training time is under an hour for 80 epochs. 

\section{Related Work}\label{sec:relatedwork}

\sn is related to a wide-range of research areas and we group the related works into key areas consisting of: (i) \nameref*{itm:rw_1}, (ii) \nameref*{itm:rw_2} and (iii) \nameref*{itm:rw_3}.

\subsection{In-situ Leaf WC Sensing}\label{itm:rw_1}

Two commonly available leaf WC sensors are \textit{(i)} The Agrihouse Leaf sensor \cite{noauthor_leaf_nodate} and \textit{(ii)} PSY1 leaf psychrometer \cite{noauthor_psy1_nodate}. The Agrihouse Leaf sensor measures leaf thickness, where the changing volume of the leaf is primarily due to water content changes. PSY1 leaf psychrometer measures leaf water potential, which is related to water stress in the plant. Both these methods however involve clamping in-situ sensors onto the leaf, which interferes with the natural photosynthesis process that is critical for plants' health.

\subsection{Remote Sensing for Leaf Water}\label{itm:rw_2}
The behavior of electromagnetic (EM) waves at boundaries is governed by a material's relative permittivity, also known as the dielectric constant. As the relative permittivity is dependent on the frequency of the EM wave, the following section is organized by frequency shown in Table~\ref{tab:emfrequencychart}.

\begin{table}[t]
\centering
  \caption{Frequency range of various EM technologies}
  \label{tab:emfrequencychart}
\begin{tabularx}{\linewidth}{l|l|X} 
\toprule
Technology & Frequency Range & Limitations \\ 
\midrule
RFID & 135 kHz – 7 GHz & require in-situ tags \\
LoRa/WiFi & 433 MHz – 5 GHz & RSSI room calibration \\
mmWave & 30 GHz – 300 GHz & directional sensitivity \\
Terahertz & 100 GHz – 3 THz & attenuation and cost\\
\bottomrule
\end{tabularx}
\end{table}

\subsubsection{RFID}

RFID tags may be placed directly on the leaf to measure leaf water status, where the change of the antenna impedance is related to the dielectric properties of the tag substrate \cite{zhang_plant_2022, dey_paper_2020}. As RFID tags are placed on relatively rigid surfaces, the main consideration when placing a tag on the leaf directly is the material the tag substrate is made from, which needs to be both flexible and lightweight. Polylactic acid is a material that has been used for leaf-based tags as it has the above properties \cite{palazzi_leaf-compatible_2019}.

\subsubsection{WiFi}

In the WiFi frequency range (2.4–5 GHz), the radio waves generally penetrate and propagate through the vegetation, with a linear correlation between plant WC and signal attenuation~\cite{536525}. This correlation has been leveraged to calculate the RWC of Eucalyptus leaves by measuring the RSS signal loss ~\cite{peden_model_2021}. In addition to plant WC, other influences on RSS signal attenuation have been explored, including factors such as leaf size, temperature and humidity~\cite{bauer_towards_2020}.

\subsubsection{mmWave}



Santos used mmWave radar to measure an orchid leaf across various events, including watering and sun/shade control in~\cite{santos_potential_2021}. While the reflected signal changed during these experiments, there was no mathematical modeling or ground truth sensor to correlate the radar signal changes to any plant physiological properties. Hoog et. al. used mass loss in the leaf as a ground truth to estimate the RWC using the radar signal amplitude in~\cite{hoog_60_2022}. Despite the use of a 2 mm Polyethylene film to mitigate leaf curling, specific sets of leaves exhibited residual measurement variability, which was attributed to variations in incident angles and surface roughness. There is still a gap in addressing the challenges of measuring an uneven leaf surface using mmWave radar.


\subsubsection{THz}

In the Terahertz frequency range, water is highly absorbed, so is an effective frequency range to measure water content in leaves \cite{gente_monitoring_2015,hadjiloucas_measurements_1999}. Spatial variability of water distribution through a leaf can easily be identified \cite{song_temporal_2018} however surface roughness has a higher impact on the reflected signal in this part of the spectrum \cite{hadjiloucas_measurements_1999}. 
    
\subsection{mmWave Sensing}\label{itm:rw_3}
mmWave Radar is currently used in a variety of industries, such as automotive and industrial applications \cite{10.1145/3372224.3419202, sarkar_deepradar_2021, zhao_m-cube_2020, wei_intelligent_2022, bae_poster_2023, bae_omniscatter_2022, king_long-range_2021}. Initially, it was used as an object detection technology but has recently been used for object classification \cite{wu_msense_2020,yeo_radarcat_2016,yeo_exploring_2018,liang_fg-liquid_2021,yang_feasibility_2019,shanbhag_contactless_2023}.
Object classification using mmWave radar works by initially calculating the distance of the target material, then calculating the reflection coefficient using the received attenuated signal. From the reflection coefficient, the relative permittivity may be calculated which allows us to identify different materials. 

One drawback of using relative permittivity for object detection is the difficulty of identifying materials with similar relative permittivity. mSense~\cite{wu_msense_2020} was able to differentiate between wood, plastic, ceramic, water and aluminum, as they have measurable differences in permittivity values. However identifying items with similar permittivity values, like mineral water and diet Pepsi, is more difficult~\cite{dhekne_liquid_2018}.

FG-LiquID~\cite{liang_fg-liquid_2021} improved classification of similar permittivity materials over previous works \cite{dhekne_liquid_2018,wu_msense_2020} by using both the signal strength and the location of the object. IndexPen~\cite{10.1145/3534601}, MilliPoint~\cite{10.1145/3432221}, Mask Does Not Matter~\cite{10.1145/3495243.3560515}, milliLoc~\cite{zhang_push_2023} and UWBMap~\cite{10.1145/3494977} also combined RSS and AoA information to improve the sensing accuracy of their mmWave systems.

\section{Conclusion}\label{sec:conclusion}

We introduce \sn, a low-cost leaf WC measurement system using a COTS mmWave radar. By utilizing Tx beam steering and Rx beamforming, \sn models the correlation between the leaf WC and radio waves by exploiting the unique interaction between the radio waves and the internal structure of leaves. Our extensive in-lab evaluation with three different leaf types demonstrates that \sn significantly outperforms prior methods, providing a highly accurate estimate of RWC with up to 62\% improvement in MAE. The live plant evaluation, for both indoor and glasshouse environments, found a strong relationship between the RWC predicted by \sn and the VWC in the soil, indicating that \sn is a reliable tool for assessing plant water stress.




\balance
\bibliographystyle{ACM-Reference-Format}
\bibliography{ch1-sensing, websites}

\end{document}